\documentclass[sigconf,edbt]{acmart-edbt2021}

\usepackage{graphicx} 
\usepackage{booktabs} 
\usepackage{array} 
\usepackage{paralist} 
\usepackage{verbatim} 
\usepackage{listings}
\usepackage{xcolor}
\usepackage{url}
\usepackage{subcaption}
\usepackage{amsmath}
\usepackage{tikz}
\usetikzlibrary{positioning, shapes.misc}

\def\ParHead{\noindent\bf}

\newcommand*\circled[1]{
  \tikz[baseline=(char.base)]{
    \node[shape=circle,draw,inner sep=1pt,fill=black,text=white] (char) {\bf{#1}};
  }
}

\newcommand\nnfootnote[1]{%
 \begingroup
  \renewcommand\thefootnote{}\footnote{#1}%
  \addtocounter{footnote}{-1}%
  \endgroup
}

\def\BibTeX{{\rm B\kern-.05em{\sc i\kern-.025em b}\kern-.08em
    T\kern-.1667em\lower.7ex\hbox{E}\kern-.125emX}}

\setcopyright{rightsretained}
\acmISBN{}
\acmYear{2022}
\title{Efficient Specialized Spreadsheet Parsing for Data Science} %

\author{Felix Henze\footnotemark{} \ \ Haralampos Gavriilidis$^{\,\diamond}$ \ \ Eleni Tzirita Zacharatou$^{\,\diamond}$ \ \ Volker Markl$^{\,\triangledown, \diamond}$} 
\affiliation{Access Microfinance Holding AG$^{\,\ast}$, DFKI GmbH$^{\,\triangledown}$, Technische Universität Berlin$^{\,\diamond}$}
\affiliation{felix.henze@accessholding.com, \{gavriilidis, eleni.tziritazacharatou, volker.markl\}@tu-berlin.de}

\renewcommand{\shortauthors}{Henze and Gavriilidis, et al.}

\begin{abstract}
Spreadsheets are widely used for data exploration. 
Since spreadsheet systems have limited capabilities, users often need to load spreadsheets to other data science environments to perform advanced analytics.
However, current approaches for spreadsheet loading suffer from either high runtime or memory usage, which hinders data exploration on commodity systems. 
To make spreasheet loading practical on commodity systems, we introduce a novel parser that minimizes memory usage by tightly coupling decompression and parsing.
Furthermore, to reduce the runtime, we introduce optimized spreadsheet-specific parsing routines and employ parallelism. 
To evaluate our approach, we implement a prototype for loading Excel spreadsheets into R environments.
Our evaluation shows that our novel approach is up to $3\times$ faster while consuming up to $40\times$ less memory than state-of-the-art approaches. 

\smallskip
\noindent \textbf{Artifact Availability:} The source code has been made available at \url{https://github.com/fhenz/SheetReader-r}.
\end{abstract}

\keywords{Data Loading, Spreadsheet Parser, Parsing Parallelization \vspace{-0.5cm}}

\begin{document}
\maketitle
\nnfootnote{$^\ast$ Work done while the author was at Technische Universität Berlin.}

\section{Introduction}
\label{sec:introduction}
Due to their intuitive layout, spreadsheets are ubiquitous for data exploration and analysis~\cite{rahman2020benchmarking,rahman2021noah}. 
While modern spreadsheet systems provide some analysis tools, such as PivotTables and aggregation formulas, they do not support more advanced tasks, such as iterative analyses and model building.
As a result, to perform their analyses, users turn to more specialized data science environments, such as R and Python, that provide ecosystems with a plethora of data science libraries.

Consider the following real-world example, which refers to a common use-case in \textit{AccessHolding}, a Germany-based microfinance investment and holding company.  
A data scientist working for a financial institution needs to determine factors indicative of default risk from loan data. 
To that end, she wants to run a logistic regression analysis.
Before deploying the logistic regression model in production, the data scientist validates it on her laptop using R.
Since the data is only available in spreadsheet files, the first preprocessing step consists of loading the data into the R runtime. 
To perform efficient analyses, users need tools that allow them to quickly load their spreadsheet data without consuming a large amount of resources.
However, although spreadsheets are widely used among data scientists, there has been little work on interoperability with data science environments. 

Typical spreadsheet applications, such as \emph{Microsoft Excel} and \emph{LibreOffice Calc}, store data as a collection of individually compressed XML structured files. 
Existing tools for converting these data collections into an appropriate data format for the target environment rely on general methods for decompression and XML parsing  (i.e., DOM and SAX)~\cite{Li2009, parsing_survey}. 
Specifically, state-of-the-art DOM-based parsers materialize the entire XML file in memory, thereby suffering from high memory usage. 
In contrast, SAX-based parsers expose a large number of parsing events through their event-based API, suffering from bad runtime performance.   

\begin{figure}
  \centering
  \includegraphics[width=.9\linewidth, keepaspectratio]{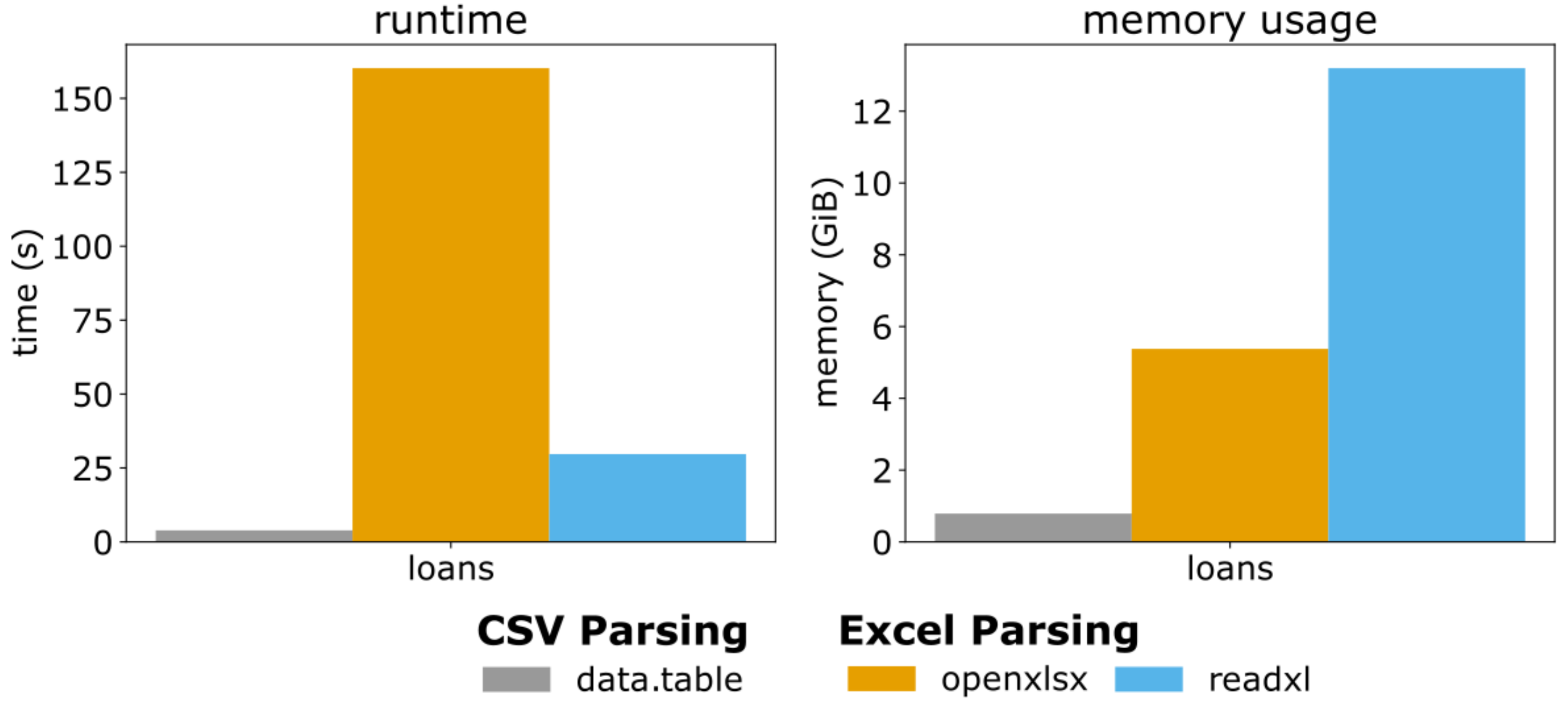}
  \caption{Performance of existing R packages for parsing a real-world spreadsheet and the corresponding CSV file.}
  \label{fig:motivation}
\end{figure}

To show the inefficiency of existing solutions, we compare two state-of-the-art Excel parsers for R (i.e., \emph{openxlsx} and \emph{readxl}) with a highly optimized CSV parser (\emph{data.table}) when loading the same data.
To that end, we use as input one real-world file in the appropriate format.  
We provide the experimental setup and configuration in Section~\ref{sec:evaluation}.
In Figure~\ref{fig:motivation}, we illustrate that the fastest Excel parser takes around 30 seconds to load 172 MB of data, while consuming up to 13 GB memory.
Compared to the CSV parser that only takes 4 seconds and consumes up to 1.1 GB memory, this is an overhead of $7.5\times$ for runtime and almost $12\times$ for memory usage.
In contrast, the most memory-efficient Excel parser consumes up to 5 GB memory, which represents an overhead of $4.5\times$, but is $40\times$ slower, taking 160 seconds to parse the file.
This gap in performance is due to the fact that spreadsheet parsers are not specialized to exploit the spreadsheet file structure.
Consequently, and given that many users work on commodity hardware (e.g., business laptops, desktops), loading spreadsheets can easily become a significant bottleneck in data science applications.

By leveraging the spreadsheet file structure and its unique properties, we overcome the aforementioned bottlenecks and enable high performance spreadsheet loading in data science environments.
We propose \emph{SheetReader}, an efficient specialized spreadsheet parser. Furthermore, we introduce two parsing approaches for SheetReader with trade-offs between runtime performance and memory consumption. 
Our first approach, \emph{consecutive} parsing, achieves very fast loading times by heavily utilizing parallelization.
The second approach, \emph{interleaved} parsing, while also employing parallelization, it primarily aims to minimize memory consumption by tightly coupling decompression and parsing. 
{\ParHead Contributions.} Our contributions are summarized as follows: 
\begin{itemize}
    \item We introduce spreadsheet-specific optimizations and employ parallelism that requires minimal synchronization to reduce the runtime for spreadsheet parsing. Furthermore, we minimize memory utilization by tightly coupling decompression and parsing.  
    \item We introduce two parsing approaches that allow users to choose between runtime and memory utilization based on their needs. The \emph{consecutive} approach achieves fast loading times through massive parallelization, but its memory utilization is data-dependent. In contrast, the \emph{interleaved} approach uses a configurable and constant amount of memory while still achieving low runtime.  
    \item We provide a general solution for different data science environments, by storing the parsed data in an environment-agnostic intermediate data structure.
    \item We experimentally show that SheetReader outperforms the existing solutions by up to $3\times$ and $40\times$ in terms of runtime and memory utilization, respectively. We also show that parallelizing the decompression further reduces the runtime by around 35\%.  
\end{itemize}

{\ParHead Outline.} Next, 
after introducing the necessary background in Section~\ref{sec:background}, we describe \emph{SheetReader}'s parsing approaches and spreadsheet-specific parsing optimizations in Sections~\ref{sec:approach}  and~\ref{sec:spreadsheet-specific_optimizations}. 
Then, Section~\ref{sec:evaluation} presents our experimental evaluation against state-of-the-art solutions using real-world and synthetic datasets. 
We discuss the related work in Section~\ref{sec:related_work} and conclude in Section~\ref{sec:conclusion}.

\section{Background}
\label{sec:background}
{\ParHead Spreadsheet Standards.}
We describe the structure of a spreadsheet, focusing on the file format currently used by Excel, known as Office Open XML (OOXML) and standardized as ECMA-376~\cite{ooxml}. 
Part 2 of ECMA-376 specifies the Open Packaging Conventions (OPC) that describes the general structure of OOXML files.
According to OPC, OOXML files are ZIP archives containing a collection of XML files. 
OPC reserves some file names and extensions for describing the types of files contained in the archive and their relationships.  

\begin{figure}[t]
  \centering
  \includegraphics[scale=0.43]{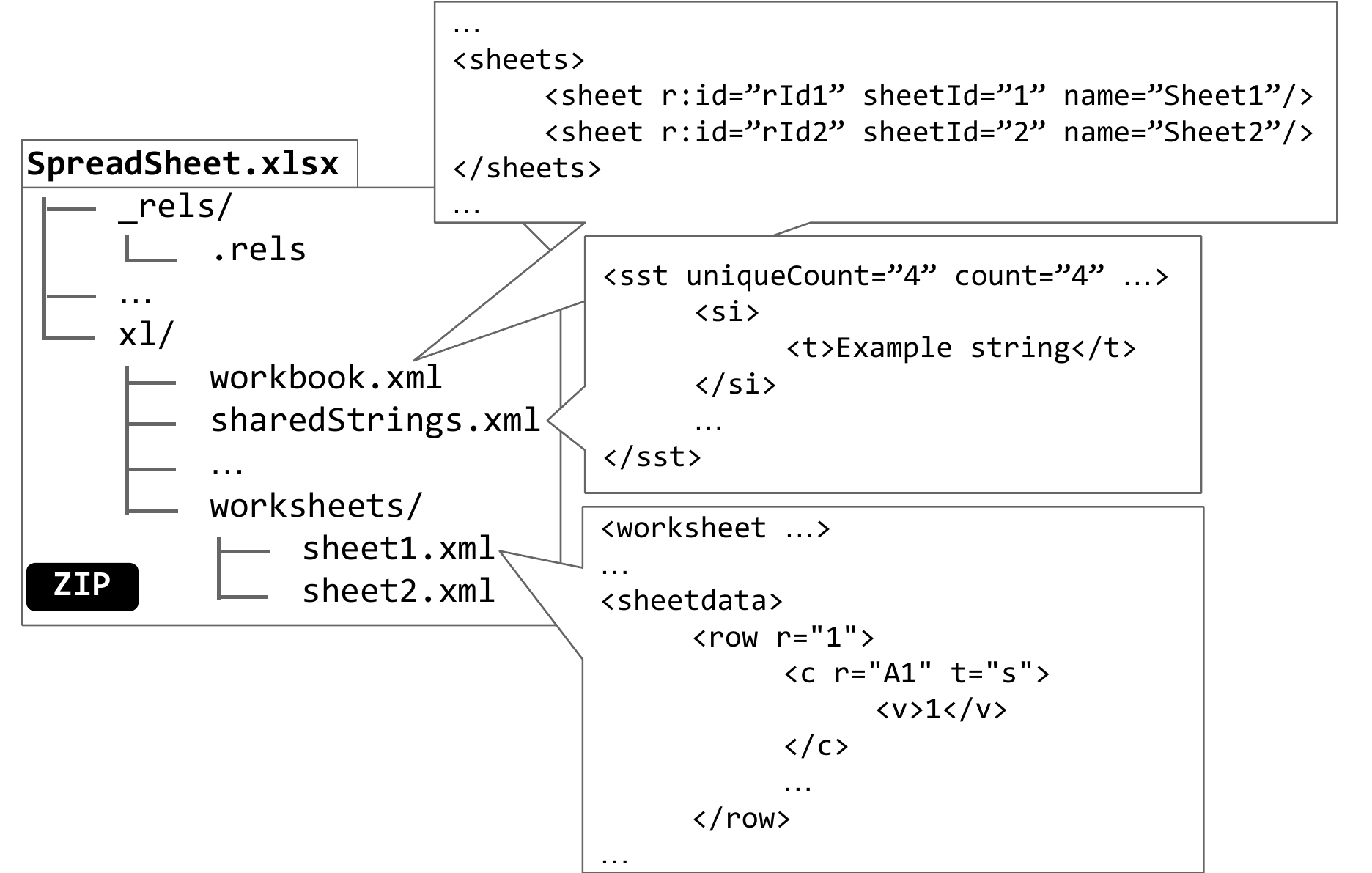}
  \caption{Spreadsheet structure (simplified).}
  \label{fig:spreadsheet_structure}
\end{figure}

\Cref{fig:spreadsheet_structure} provides a simplified overview of the spreadsheet structure.
Excel documents consist of a workbook that can contain several worksheets.
The workbook determines the names, IDs, and archive locations of all the spreadsheets.
The worksheets, e.g., \verb+sheet1.xml+ in the figure, store the actual data.
Additionally, Excel saves strings in a separate file from the actual worksheets, \verb+sharedStrings.xml+,  where they are referenced by index.
The top level reserved files contain metadata that allows to identify files relevant for further processing and serve as an entry point for programs. 
Specifically, valid Excel files require the top level relationship file \verb+/_rels/.rels+ that specifies the locations of the workbook and the shared strings file inside the archive.

As stated in the OOXML specification, the XML files in the ZIP archive can be either uncompressed or compressed using the \texttt{Deflate} format.
\texttt{Deflate}~\cite{deflate} is a block-based compression format with dynamic block sizes.
It arranges the blocks in a stream and compresses them individually.
Although it is possible to compress an entire document into a single large block, using smaller blocks typically improves the compression ratio.
Within a block, \texttt{Deflate} uses duplicate string elimination, a technique where duplicate series of byte streams are replaced with back-references to the previous identical byte stream, which can in turn also be a back-reference.
Back-references can point to previous blocks, as long as the distance does not exceed a sliding window of the last 32 KB of decompressed data.
As a result, \texttt{Deflate} documents are challenging to decompress in parallel, because to decompress a given block, all previous blocks need to be decompressed first.

{\ParHead XML Parsing.}
There exist two dominant approaches for XML parsing, DOM (Document Object Model) and SAX (Simple API for XML)~\cite{Li2009, parsing_survey}.
The DOM approach maps the XML file contents to an in-memory tree and provides an interface that allows to In contrast, the SAX approach exposes an event handling interface.
While traversing the XML document, the SAX parser fires events for the found XML entities (tags), which then trigger the previously registered handlers.
DOM is well-suited for random access.
However, a major disadvantage regarding resource consumption is that it needs to materialize the whole document in memory before parsing it.
SAX parsers do not experience this bottleneck.
However, the event handling interface makes it challenging to keep track of the entire document while parsing it, and leads to inefficient implementations.

Even though DOM and SAX approaches provide a solution for parsing XML documents, they are both very generic, i.e., \emph{they are designed to support arbitrary XML structures}.
We observe that for parsing spreadsheets it is not necessary to employ such generic approaches, as spreadsheet XML files have a very  specific XML file structure that is defined by their specification.  
We argue that a \emph{specialized parser for spreadsheets} can exploit their specific structure and find the sweet spot between DOM and SAX approaches, thereby offering reasonable memory consumption and fast runtime performance at the same time.

\section{SheetReader}
\label{sec:approach}
In the following, 
we give an overview of SheetReader's architecture (Section~\ref{sec:design}).
Then, we describe in detail our two specialized parsing approaches (Section~\ref{sec:design_parsing_methods}). 

\subsection{SheetReader Overview}
\label{sec:design}
\begin{figure}[t]
  \centering
  \includegraphics[width=0.45\textwidth]{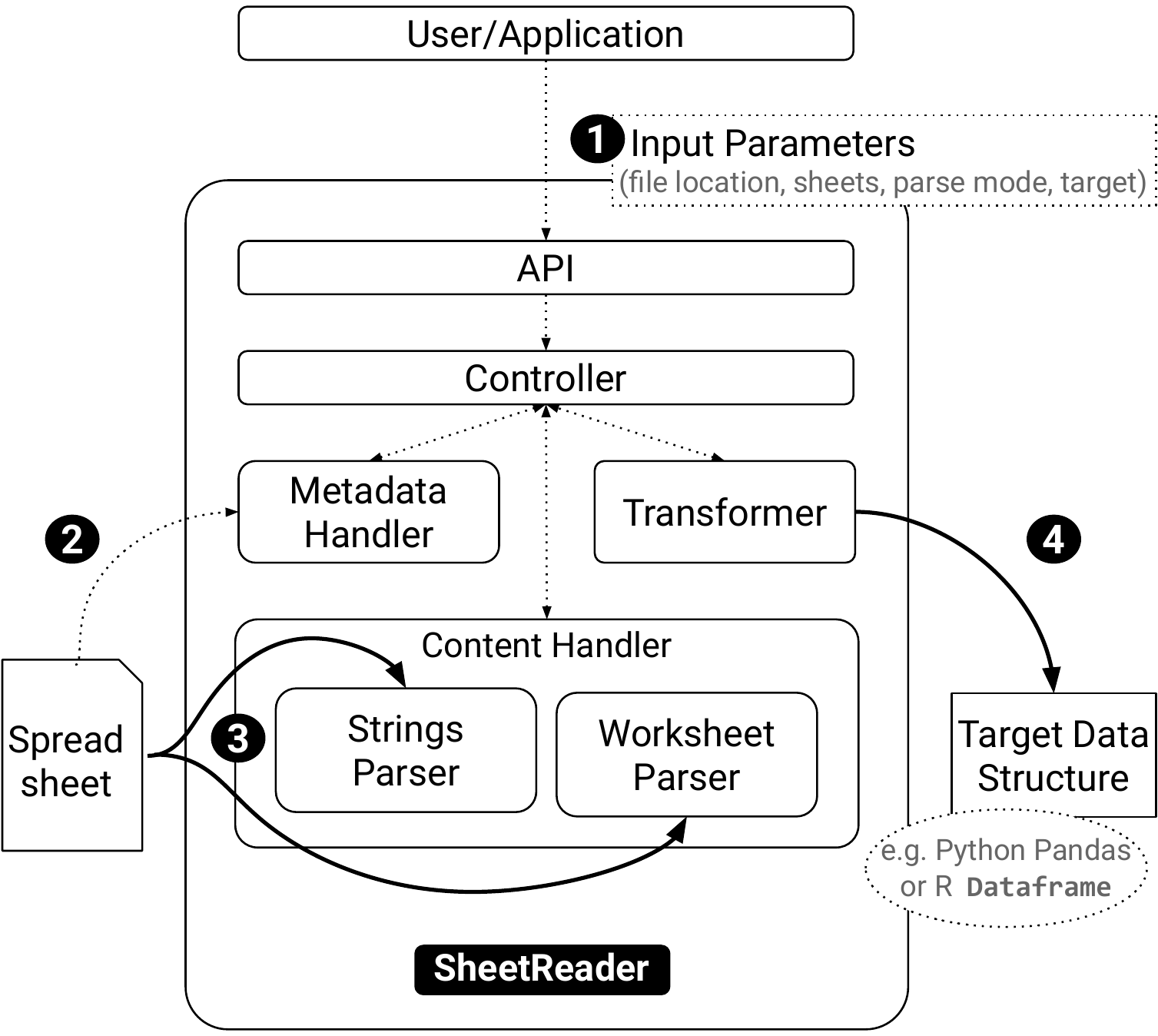}
  \caption{SheetReader's Architecture.}
  \label{fig:architecture_overview}
\end{figure}

We show an overview of our approach in Figure~\ref{fig:architecture_overview}.
SheetReader expects as input parameters related to a spreadsheet file, and loads the worksheet contents into a data structure within the target environment.
Users and applications submit parsing requests to SheetReader through its \textbf{API}, by providing I/O and parser configuration parameters~\circled{1}.
The \textbf{Controller} is the core component responsible for coordinating the overall loading routine.
At first, the Controller fetches worksheet metadata~\circled{2}, e.g. file location and sheet names, through the \textbf{Metadata Handler}.
Then, the Controller initiates the loading routine by providing the sheet names and parse mode to the \textbf{Content Handler}, a component that decompresses input files and parses the spreadsheet content into an intermediate data structure~\circled{3}.
The Content Handler has two modules, the \textbf{Strings Parser}, which is responsible for parsing the shared strings XML file, and the \textbf{Worksheet Parser}, which is responsible for parsing the worksheets containing numeric data.
These two parsers may operate in parallel, and have two different parsing modes that we describe in \Cref{sec:design_parsing_methods}.
To avoid costly reallocations for resizing the intermediate data structure during parsing, the Controller pre-allocates memory by relying on the available metadata, such as the file offset, archive size, and total strings number in the shared strings file.
Our parsers assume valid spreadsheets as input, since spreadsheet systems, e.g. Excel, are unlikely to produce corrupt files.

When parsing is completed, the \textbf{Transformer} executes the final loading step, i.e., creating the target data structure \circled{4}.
Contrary to the worksheet, SheetReader stores the cell data in column-wise data layout.
This allows to transform intermediate data to column-based target data structures widely found in data science environments, e.g., R and Python Pandas \texttt{Dataframe}s, without reconverting the layout.
Additionally, SheetReader's internal intermediate data structure enables to reuse its core parsing routines in different runtimes by only implementing the Transformer interface and language bindings.
This interface exposes methods for transforming the intermediate data structure into a target data structure.
For example, our prototype Transformer implementation in R converts intermediate data into a \texttt{DataFrame}.

\subsection{Spreadsheet Parsing Approaches}
\label{sec:design_parsing_methods}
We introduce two parsing approaches that both the Strings and the Worksheet Parser components can use.
These approaches express a trade-off between runtime performance and memory efficiency.
The \emph{consecutive} approach is optimized for runtime performance and the \emph{interleaved} one for minimal memory usage.
Users can choose their preferred approach based on their needs.

Both approaches rely on the same general parsing routine that we outline here.
As our parser targets specialized XML documents, it operates by finding the opening and closing tags for specific XML elements.
For example, as shown in ~\Cref{fig:spreadsheet_structure}, in Excel files, cell values \texttt{<v>val</v>} are enclosed in \texttt{<c></c>} tags, where the character sequence \texttt{<c\textvisiblespace} indicates the opening tag for a new cell.
Inside this tag, there are attributes that contain cell metadata.  
Attributes are name-value pairs that are linked through the \texttt{=} character and are separated from other pairs by a whitespace.
We parse this metadata as it defines the cell location and type, which we use to determine where to store the cell data in our intermediate data structure.
The character \texttt{>} denotes the end of the cell opening tag. 
Inside the \emph{c} element, we look for the \texttt{<v>} opening tag that contains the cell value.
We parse the value until we encounter the closing tag \texttt{</v>} and insert it into our intermediate data structure.

\begin{figure}[t]
  \centering
  \includegraphics[scale=0.43]{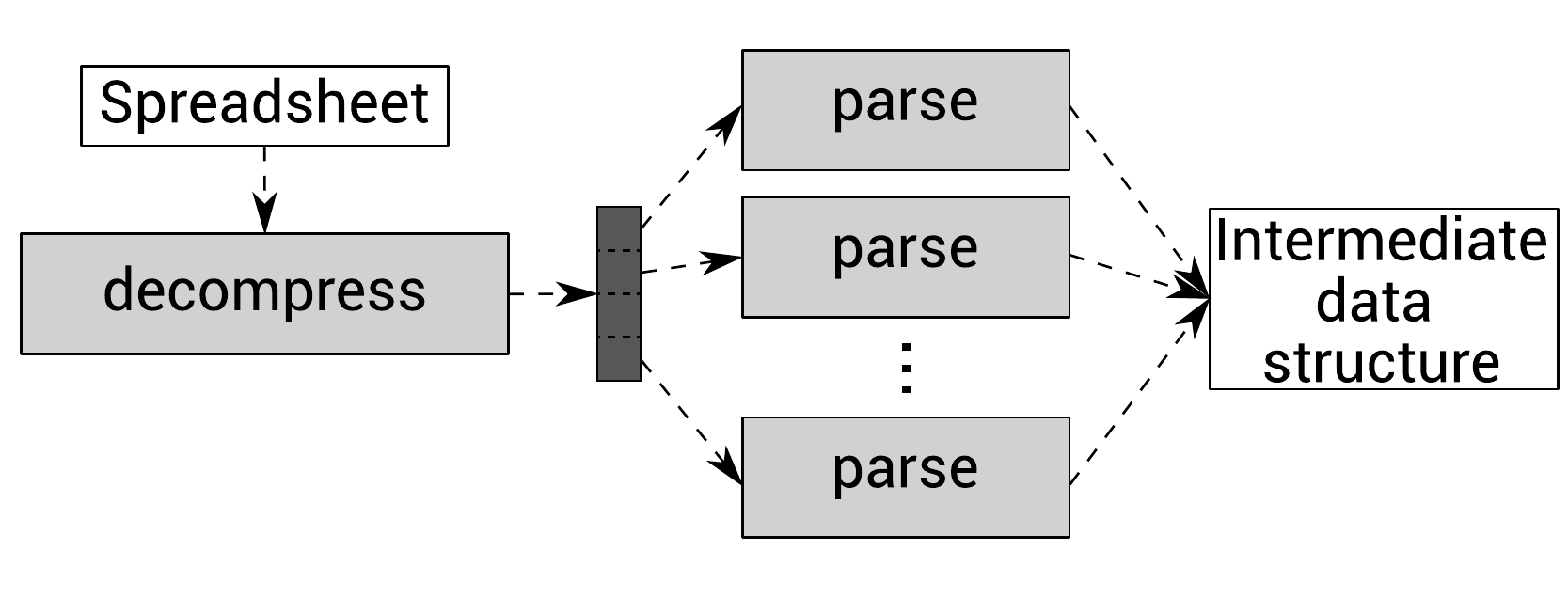}
  \caption{Consecutive Parsing.}
  \label{fig:consecutive_overview}
\end{figure}

\subsubsection{Consecutive Parsing}
We optimize \emph{consecutive} parsing for runtime performance.
As shown in Figure~\ref{fig:consecutive_overview}, \emph{consecutive} parsing first decompresses the complete document into memory and then parses the content with multiple parallel parsing threads. 
Having the complete document in memory during parsing has several advantages. 
First, we do not need to use intermediate buffers to store values for later parsing because the document itself serves as a buffer.  
This reduces costly memory operations such as allocations and copies. 
Additionally, since decompression is independent of parsing, the choice of decompression method is flexible, and we can use libraries that are optimized for full-buffer decompression. 
However, keeping the entire document in memory during parsing leads to inflated memory usage.
We note that if the document cannot fit in memory, then SheetReader uses \emph{interleaved} parsing instead.

Once the entire document has been decompressed, we can parallelize the parsing by simply splitting the document into roughly equal-sized chunks and processing each chunk by a separate thread.
However, splitting XML documents into multiple chunks that can be parsed in parallel is a challenging problem~\cite{lu2006parallel,pan2007static,pan2008transducers}. 
In particular, a parser that starts at an arbitrary point in the document lacks the context to determine how to process the encountered characters. 
To overcome this problem, we leverage the fact that spreadsheets have a predefined XML structure and determine the parsing state by identifying the type of the first XML element that we encounter in the chunk. 
Specifically, we scan the chunk for structural characters that denote the start or end of an XML tag (e.g., \texttt{<}).
We then build additional context by determining the type of the corresponding XML element.
For example, if we encounter the opening tag to a row element, we know that we are at the beginning of a new row, while if we encounter the closing tag to a cell element, we know that afterwards there is either another cell or the end of the row.
The above approach is possible because the encoding of structural characters is different when these characters are not structurally significant, e.g., when they are part of an element or an attribute value. 
For example, while \texttt{<} denotes a structural character, the same character is encoded as \texttt{\&lt;} inside an element or an attribute value.  

Our parallel parsing approach also assumes that each cell has information about its location (i.e., the row and column number), so that individual threads can determine where to insert the parsed values in SheetReader's intermediate data structure.
Although this information is not part of the standard, the most widely used tool, Microsoft Excel, provides it.
If there is no location information, we can employ an additional processing step, either before or after the parallel parsing.
Before parsing, we can perform a reduced sequential scan over the document to count the rows and cells and calculate the offset for each chunk.
This sequential scan can be implemented efficiently such that it does not significantly affect the runtime. 
An alternative approach is to let each thread insert the parsed values into its own separate intermediate data structure.
In the last step, i.e., when converting to the target, we can then merge the partial data structures by sequentially retrieving the values.

Additionally, we determine the size of the worksheet, i.e., the number of rows and columns, from the dimension element in the spreadsheet document metadata.
If the dimension element does not exist, and since we have the entire uncompressed document in memory, we can also determine the size by examining the row and column number of the last cell. %
Predetermining the worksheet size allows to pre-allocate the intermediate data structure and avoid costly resize operations.
Furthermore, it enables multiple threads to insert values in the data structure without any write synchronization mechanism.
Being unable to pre-allocate the intermediate data structure adds only minor complexity. 
When a column becomes full, we simply need to allocate a larger amount of space and copy over the existing values.
During the resizing operation, we also need a synchronization mechanism (e.g., a lock) that blocks the insertion of new values. %

Overall, each parsing thread of the \emph{consecutive} approach takes as input the starting offset of its chunk and the end offset or the total chunk length. 
Then, it locates the first cell in the chunk as discussed previously and proceeds with parsing from there, skipping over all content before the first cell.
This skipped content is actually relevant to the last cell of the previous chunk. 
Therefore, to ensure that all elements will be parsed, each thread finishes parsing the last cell of the chunk by extending its assigned parsing area over the beginning of the following chunk.

\begin{figure}[t]
  \centering
  \includegraphics[scale=0.43]{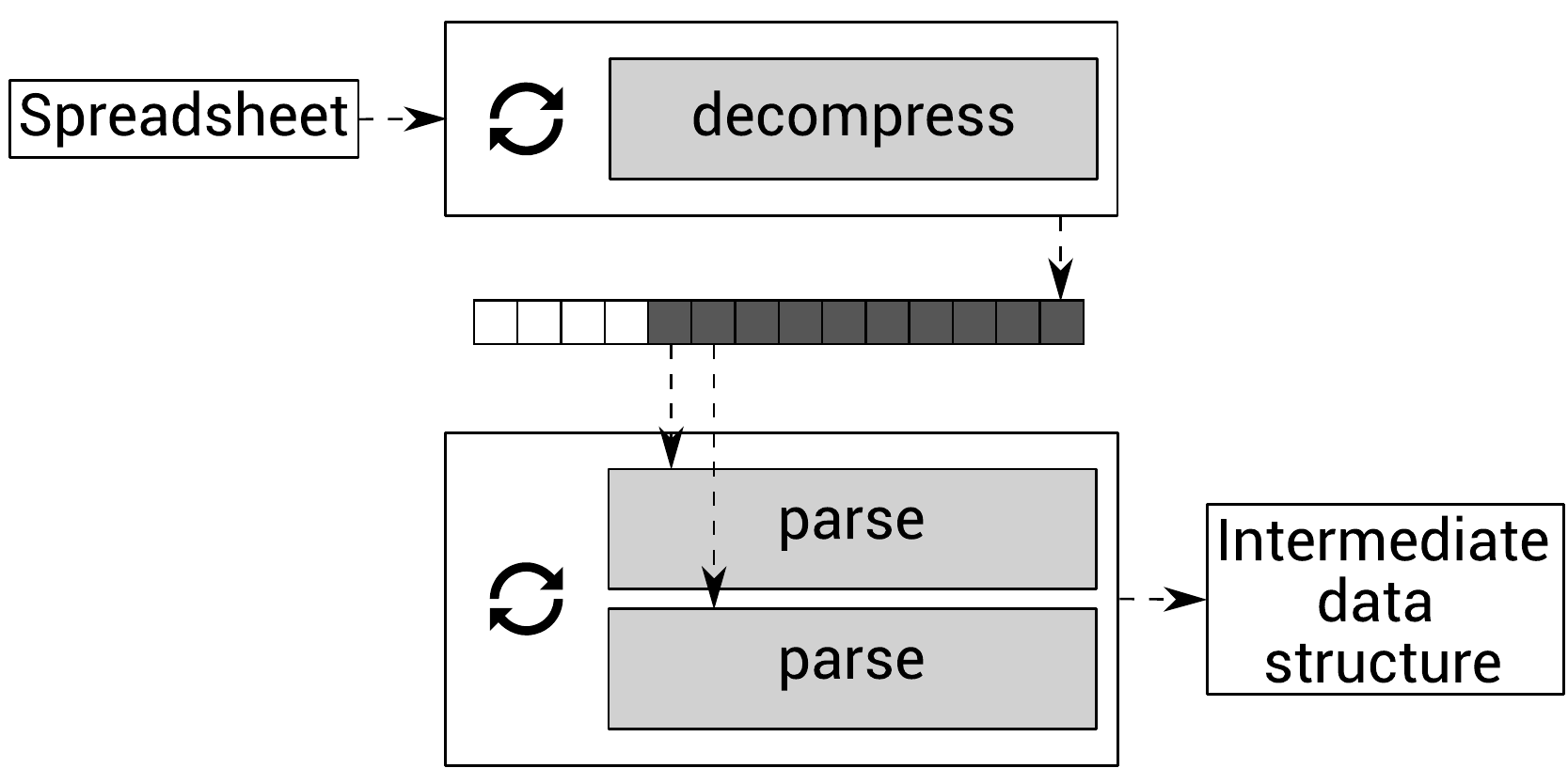}
  \caption{Interleaved Parsing.}
  \label{fig:interleaved_overview}
\end{figure}

\subsubsection{Interleaved Parsing}
This approach aims to minimize memory usage. 
To that end, it continuously recycles a constant amount of memory so that the memory usage is independent of the input document, and interleaves decompression and parsing.
Specifically, as depicted in Figure~\ref{fig:interleaved_overview}, decompression and parsing occur repeatedly one after the other. 
First, the decompression stage decompresses part of the document.
Then, the parsing stage processes this part and returns the control flow to the decompression stage, waiting for the next part to be decompressed.
As a result, it is impossible to access arbitrary parts of the document at any time and the parser is unable to backtrack or look ahead very far in the document. 
This means that the parser needs to process any relevant content as soon as it encounters it, or store it immediately for later processing.
Consequently, \emph{interleaved} parsing imposes more restrictions on the employed decompression and parsing techniques compared to \emph{consecutive} parsing.

To implement a single-threaded version of the interleaved approach, we only need a single-element shared memory buffer.
The decompression stage fills the buffer with decompressed content, and the parsing stage parses it.
However, to enable parallelization, we need a buffer with at least two elements.
Using multiple elements allows to decouple the decompression and parsing stages and execute them in parallel by separate threads.
The decompression thread writes the elements that are available for writing, while the parsing thread reads from the written elements and subsequently re-enables them for writing. 
Using a two element buffer, the threads can switch their elements only when they both finish processing. 
Since the decompression and parsing time are data-dependent, the time that a thread has to wait for the other thread to finish can fluctuate significantly. 
To reduce the total wait time and mitigate the resulting unpredictable runtime, our microbenchmarks showed that it is better to use a larger buffer.

\begin{figure}[t]
  \centering
  \includegraphics[scale=0.2]{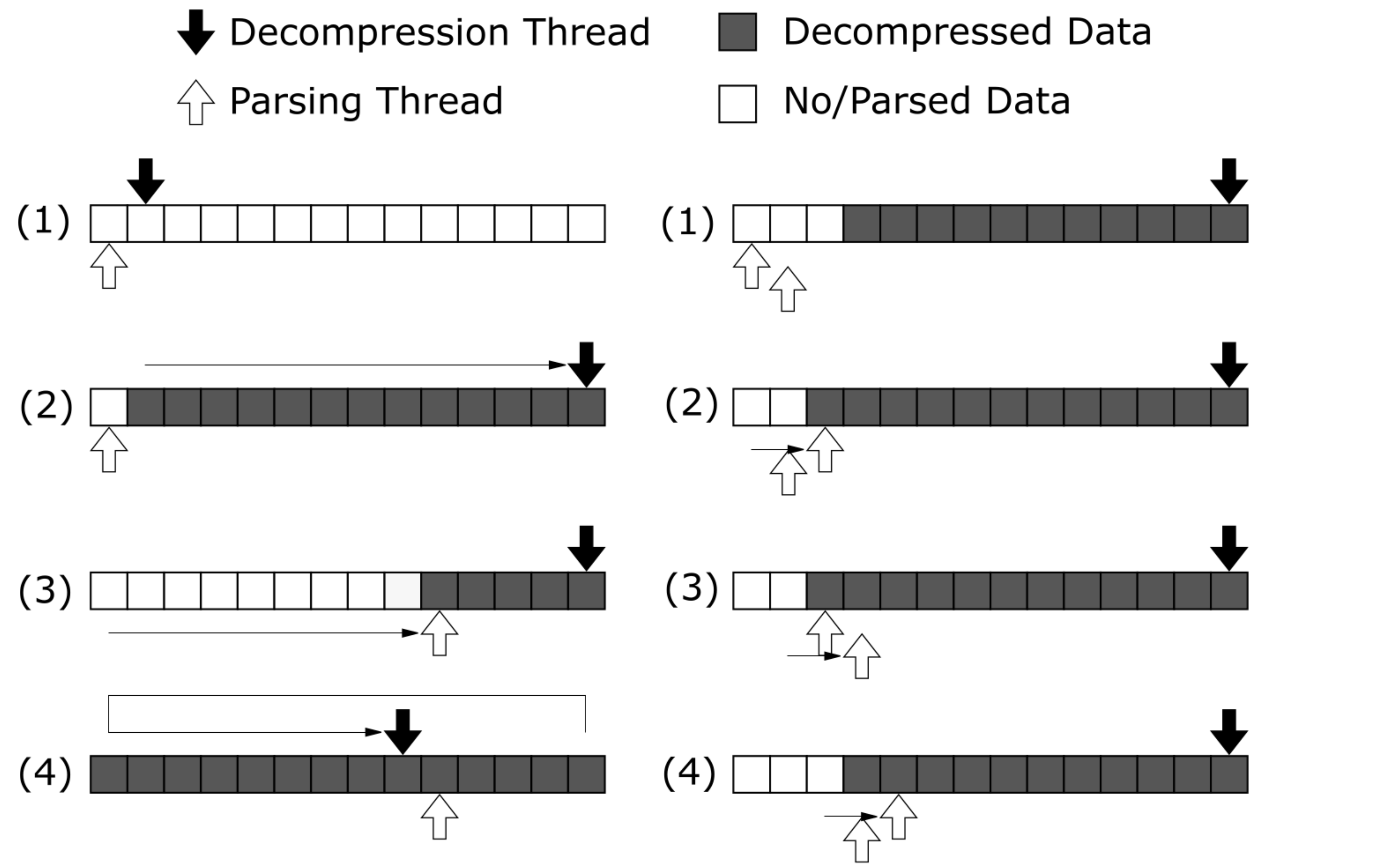}
  \caption{Concurrency Control. 
 }
 \vspace{-0.1cm}
  \label{fig:pipeline_buffers}
\end{figure}

Figure~\ref{fig:pipeline_buffers} (left) shows how the \emph{interleaved} approach works with a \emph{circular buffer} using a single parsing thread. 
The decompression and parsing threads iterate through the elements sequentially, i.e., the decompression thread writes its results into the first element, then the second one, and so on (step 2), while the parsing thread reads the elements in the same order (step 3).
When a thread reaches the last element, it loops back to the first one (step 4).

To prevent the threads from using an invalid element, i.e., the decompression thread overwriting an element that is not parsed yet or the parsing thread parsing an element that is not written yet, we make the buffer \emph{thread-safe}. 
Specifically, we use an index that indicates the element that each thread is currently operating on, and ensure that the parsing thread remains at least a single element behind the decompression thread.
That is, if the decompression thread is currently writing into the element with index $x$, the parsing thread can process all elements with up to and excluding index $x$.
If the parsing thread reaches this point, it simply blocks until the decompression index advances.
For simplicity, the initial state satisfies this requirement by starting the decompression thread one element ahead of the parsing thread (step 1). 
To ensure that the parsing thread processes the last element, we increment the write index by one when the decompression thread finishes the current round.
The decompression thread determines if it is allowed to write to an element by simply checking if that element is currently being parsed by the parsing thread.
This can happen only when the decompression thread has filled all available elements and the parsing thread has not freed any element yet, as shown in step 4.
We store the indexes as atomic integers, so that all threads see the same value when they access an index simultaneously.

In addition to parallelizing decompression and parsing, we also parallelize the parsing stage, i.e., use multiple parsing threads as shown in Figure~\ref{fig:pipeline_buffers} (right). 
We explore this avenue since our preliminary benchmarks showed that decompression is typically faster than parsing. 
The \emph{interleaved} approach can be easily extended to support parallelism in the parsing stage.
Contrary to \emph{consecutive} parsing, the parsing threads do not work with a large buffer containing the entire document, but with small buffer elements that contain only small parts of the document. 
As a result, the mechanism for distributing the elements among the parsing threads is slightly more complex.
One solution would be to introduce a flag for each element that indicates its state, i.e., if the element is ready to be parsed, ready to be written, or currently being processed. 
The threads would then pick an element to process based on these flags.
Instead, we decided to extend our existing index-based synchronization mechanism. 
Specifically, each individual parsing thread has a separate index and checks up to which element it is allowed to parse.
The decompression thread simply checks if any of the parsing threads works on the element where it wants to move next.

One remaining issue is preventing the parsing threads from parsing an element multiple times, i.e., uniquely assigning elements to parsing threads. 
This is achieved by initializing their indexes in a staggered manner and advancing them by the number of parsing threads rather than singular increments as shown in Figure~\ref{fig:pipeline_buffers} (right). 
For example, with three parsing threads, the first one starts at index 0, advancing to 3 and then 6. 
The second thread starts at index 1 and advances to 4 and then 7. 
The third one starts at index 2 and advances to 5 and then 8. 
This approach guarantees that all elements are fully processed exactly once.

Since we process the elements in sequential order, we also process the document sequentially, which enables using the parsing "extension" mechanism described for the \emph{consecutive} approach. 
If a parsing thread reaches the end of its assigned element but the last cell has not been fully parsed yet, it simply advances into the next element to finish parsing. 
Afterwards, it readjusts its index to prevent overlap with the other threads. 
This is possible because the decompression thread only writes to elements up to the last parsing thread, i.e., every element in front of a parsing thread up to the decompression thread will always be valid.

Similarly to the \emph{consecutive} approach, we exploit the predefined XML structure to deduct the parse states.
However, dealing with the lack of location information is harder because the parsing positions are constantly changing.
Each time a parsing thread advances, it skips over potentially multiple elements that contain the logical continuation of its acquired parse state. 
As a result, the parsing threads are repeatedly placed in unknown and ambiguous parse states.
We can adapt both solutions that we discussed for the \emph{consecutive} approach here.
Before the actual parsing, each parsing thread could perform a fast reduced scan over its assigned element to count the number of contained rows and cells, accounting also for the location information after blank cells.
Then, all threads would need to share their results to determine the row and cell numbers at the beginning of all elements.
Afterwards, the parsing threads would proceed with the actual processing.
Alternatively, we could create an intermediate data structure for each element rather than for each thread.
Finally, if we are unable to pre-allocate the intermediate data structure, we can apply the same solution of simply synchronizing the write and resize operations as in the \emph{consecutive} approach.

\section{Optimizations for Spreadsheets}
\label{sec:spreadsheet-specific_optimizations}
Aside from parallelization, we employ some further spreadsheet-specific optimizations to accelerate parsing, thereby further reducing the runtime.
These optimizations aim to reduce the amount of work per input character.
Ideally, when a character does not provide any relevant information, we do not want to perform any work for it. 
Additionally, we do not want to visit any given character, including potential copies of it, more than once.

Our first optimization consists of parsing element names on-the-fly rather than copying the encountered characters into a new buffer and comparing against the complete string.
We achieve this by checking if  the scanned input characters match any of the predefined known element names. 
Taking the row element as an example, we add an integer field to the parsing state that checks if the current element name is \texttt{row}.
At the start of parsing, we initialize this field to 0.
If the parser is in the appropriate state and encounters an \texttt{r} character, we increment the field. 
If we encounter an \texttt{o} right after, we increment the field once more. 
We apply the same procedure for the \texttt{w} character. 
If at any point we encounter a different character than expected, the field is reset to 0. 
Upon encountering a whitespace character, we simply determine the currently parsed element by checking the integer fields of the relevant element names. 
If the field matches the length of the checked element name (e.g., 9 for \texttt{sheetData}, 3 for \texttt{row}, 1 for \texttt{c}), this means that we just encountered the corresponding element.

\begin{figure}[t]
  \centering
  \includegraphics[width=0.8\linewidth]{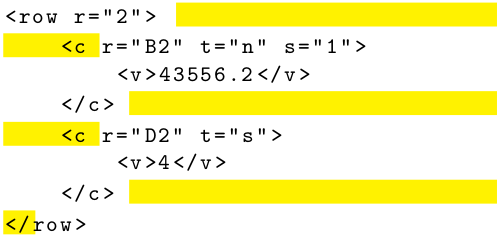}
  \caption{Excel worksheet XML extract. The highlighted sections are scanned for the opening tag of a cell element.}
  \label{fig:excel_xml_extract}
\end{figure}

Our second optimization consists of skipping as much unneeded content as possible while also determining when to skip as quickly as possible. 
In other words, we aim to determine as early as possible the amount of required work for a character and then only perform this required work.
We can identify opportunities to apply this optimization by examining the XML schema that is given by the specification.
Using the Excel format as an example, we need to check for the opening tag of a cell element (\texttt{c}) only when we have encountered a row (\texttt{row}) opening tag previously and have not encountered a row closing tag since then (cf. Figure~\ref{fig:excel_xml_extract}). 
This also applies for values (\texttt{v}) inside cells (\texttt{c}), rows (\texttt{row}) inside the sheet data (\texttt{sheetData}), and even for locating the sheet data element itself.

Furthermore, we avoid parsing and deserializing attributes that do not contain relevant data or metadata for creating the target data structure.
For example, all row elements in Excel worksheets have an attribute that indicates the height of the row.
Such irrelevant tags and their values should be skipped as early as possible.
Given the XML format, we achieve this by skipping all content between the opening and closing quotation marks of the irrelevant attribute value. 
We note that we assume that the input document is a valid XML conforming to the specification.
Otherwise, if the XML contains invalid values, e.g., if a quotation mark is missing, the parser might skip some relevant data.

To avoid visiting characters more than once, we try to perform parsing in-situ without using any intermediate copies.
This is particularly relevant for the \emph{interleaved} parsing approach, where there are no guarantees regarding which part of the document is currently available in memory.  
For example, the value of the row number attribute might be split between buffer elements. 
Since it is impossible to access the first element once we advance to the second one, a naive solution would copy the relevant portion from each element into another intermediate buffer, so that the two parts can be combined and the complete value can be deserialized.
We avoid such copies, and thus processing the same character twice, by deserializing the characters as they arrive.

Deserializing integers in-situ is simple. 
We first initialize the integer value to 0 and then for every read character, we multiply the current value by 10 and add to it the deserialized character.
We can use this approach to deserialize most of the required attribute and element values from the worksheet.
We can, for example, use a virtually identical mechanism for spreadsheet form numbers where "A" corresponds to 1 and "AA" to 27.
The only difference is that we need to multiply by 26 and adjust the deserialization of the characters to numbers.
Other values such as booleans or cell types are also trivial to deserialize without copying.
However, we cannot apply the above technique to deserialize floating point values in-situ, as it can potentially introduce rounding errors and thus produce erroneous results.   
Thus, for floating point values, we cannot avoid copy buffers. 

Overall, our spreadsheet-specific optimizations improve the performance of the low-level parsing routine. %
As a result, in single-threaded execution, the optimizations directly translate into lower runtime.
In the case of multiple threads, the optimizations accelerate the execution of each individual thread.
Consequently, we can use fewer threads, thereby potentially reducing synchronization overheads.

\section{Evaluation}
\label{sec:evaluation}
In this section, we first describe the experimental setup and methodology and then present a thorough evaluation of \emph{SheetReader} in terms of runtime and memory usage.
To demonstrate \emph{SheetReader}'s benefits, we first compare it with existing state-of-the-art solutions for spreadsheet parsing and then perform an in-depth analysis to study the trade-offs between our proposed parsing approaches.
Lastly, we evaluate parallel decompression to determine its impact on the runtime.

\subsection{Experimental Setup \& Methodology}
\vspace{0.1cm}
{\ParHead Hardware Configuration.} 
The experiments were performed on a machine equipped with an AMD EPYC 7702P 64-Core CPU, 512 GB RAM, and a 512GB SSD, running Ubuntu 20.04 (kernel version 5.4.0-90).

{\ParHead Benchmarks.} We use various benchmarks to measure the runtime and memory usage of our approach and the competing ones. 
Following our prototype, the benchmarks involve loading an Excel spreadsheet file into R. 
As our prototype targets the xlsx format introduced in 2007, it is impossible to execute benchmarks designed for older format versions.
We run every benchmark on a new R instance to avoid potential residual objects in memory from influencing later measurements. 
While the instance is running, we periodically measure its memory usage. 
For the general comparison between the approaches, we use the maximum measured memory usage and the total runtime.
Additionally, we insert logging messages that indicate the beginning and end of individual loading stages.
This periodic data allows us to examine individual benchmarks in detail.
That is, we associate the separate steps of each approach with particular messages and determine the impact of each step on the overall memory usage.
We repeat each benchmark $5$ times and report the average.
We assume cold system caches, i.e., we clear OS caches before re-executing each benchmark. 

{\ParHead Datasets.} Most of our benchmarks use synthetically generated Excel spreadsheets according to specific desired parameters such as the percentage of numeric vs. text values or the percentage of blank cells. 
We generate Excel spreadsheet files for various row counts where larger spreadsheets are supersets of smaller ones.
The compressed sizes range from 13.6 MB (10,000 rows), to 413 MB (300,000 rows), up to 827 MB (600,000 rows).  
Unless otherwise specified, each spreadsheet has $100$ columns and contains only numeric values, without any blank cells.
Furthermore, we use two real-world financial spreadsheet files from \textit{AccessHolding} to study the performance of our parser in comparison with the state-of-the-art in a real setting.
For data protection reasons, we anonymized the files before running our benchmarks.
The first file, \emph{loans}, has $280,973$ rows, $110$ columns, and a compressed size of $172$ MB.
The second file, \emph{transactions}, has $447,241$ rows, $84$ columns, and a compressed size of $193$ MB. 
Both files contain a mix of different data types, i.e., integers, dates, floats, booleans, and text.
While the first file has only a few empty cells, the second one has significantly more (i.e. ~20 columns are almost empty).

{\ParHead Baselines.} 
We chose to implement our prototype in R because of its popularity among data scientists.
Hence, we experimentally compare \emph{SheetReader} with existing R packages for loading spreadsheets.
After analyzing several packages for Excel parsing, we chose to include the \texttt{openxlsx} and \texttt{readxl} packages\footnote{\url{https://github.com/ycphs/openxlsx},  \url{https://readxl.tidyverse.org/}} as they showed the best performance for lowest memory usage and fastest runtime, respectively.
Both packages work solely with Excel files and are written in C\texttt{++}. 
\texttt{Openxlsx} employs a hybrid DOM/SAX approach, and extracts cell values using regular expressions, while \texttt{readxl} first constructs a DOM tree from the spreadsheet XML using the XML DOM parsing library \texttt{RapidXML} and then processes the tree further to extract the cell values.

{\ParHead Software Configuration.} We use the following versions of R, R packages, and libraries: \texttt{R} {4.0.3}, \texttt{data.table} {1.13.2}, \texttt{openxlsx} {4.2.3}, \texttt{readxl} {1.3.1}, \texttt{miniz} {2.1.0}, \texttt{libdeflate} {1.7}. 
We implemented our prototype in C\texttt{++} and compiled with \texttt{gcc/g++} {9.3.0}.
By default, decompression uses $1$ thread and parsing uses $8$ and $2$ threads for the \emph{consecutive} and the \emph{interleaved} approach, respectively. 
If applicable, shared strings are parsed in parallel using one additional thread.
In the \emph{consecutive} approach, we determine the buffer size for the decompressed content from the ZIP metadata.
In the \emph{interleaved} approach, we allocate a buffer with 1024 32KB-sized elements after empirically evaluating several configurations.

\subsection{Comparative Analysis}

\begin{figure}
  \centering
  \includegraphics[width=\linewidth, keepaspectratio]{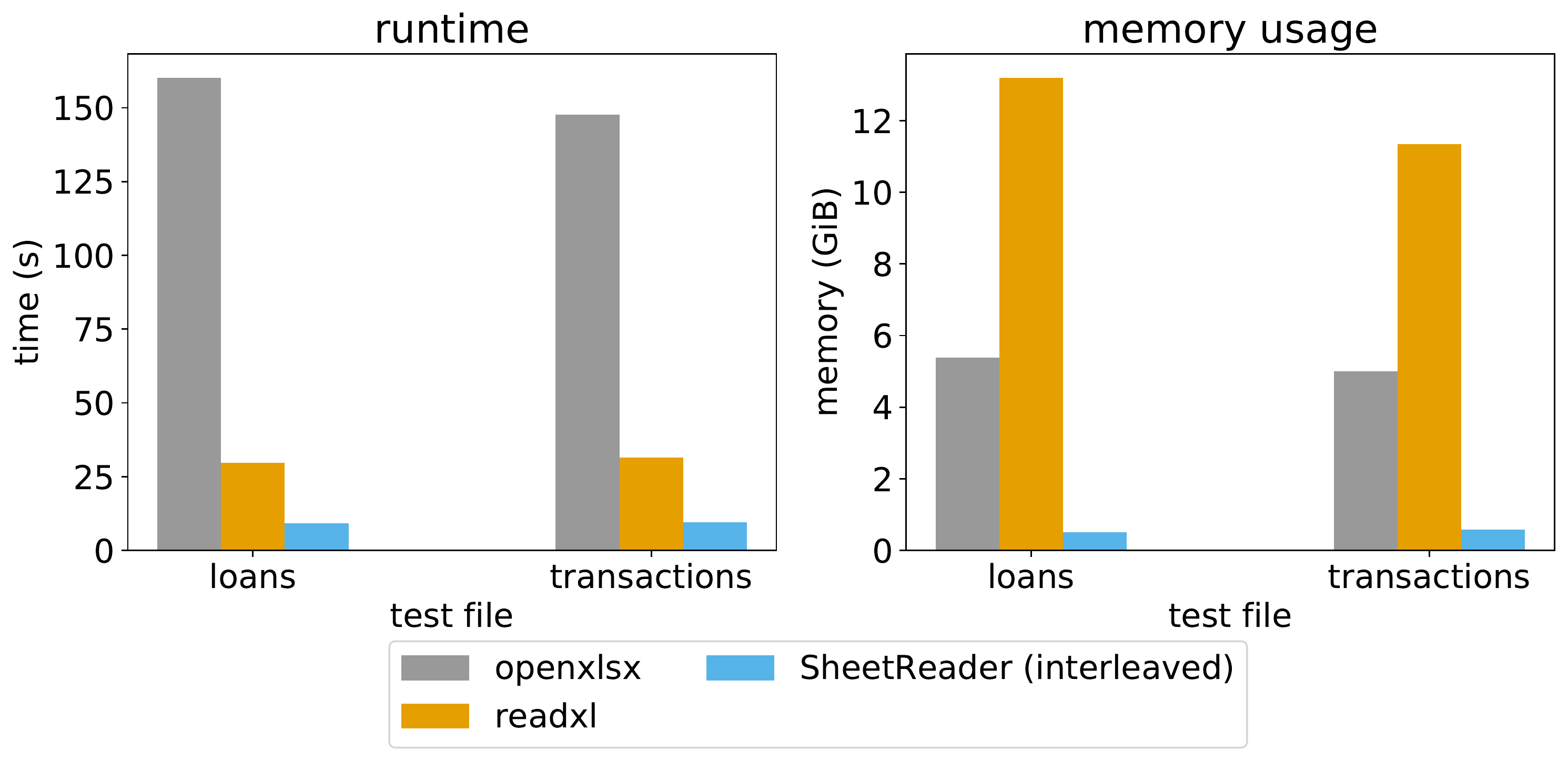}
  \caption{Performance overview of \emph{SheetReader} and existing R packages for parsing real-world spreadsheets.}
  \label{fig:evaluation_overview_real}
\end{figure}

{\ParHead Comparison using Real Data.} 
\Cref{fig:evaluation_overview_real} compares \emph{SheetReader's} \emph{interleaved} parsing approach with the state-of-the-art in terms of runtime and memory usage on the two real-world datasets. 
Furthermore, we parse the strings parallel to the worksheet in a single thread that performs both decompression and parsing. 
\emph{SheetReader} is $3.2\times$ faster than \texttt{readxl}, the fastest existing solution, while also consuming $26\times$ and $20\times$ less memory for loans and transactions, respectively.  
Compared to the most memory-efficient existing solution, \texttt{openxlsx}, \emph{SheetReader} is $17\times$ faster with $10.7\times$ less memory consumption in the case of the loans file.
In the case of transactions, it is $15\times$ faster with $8.6\times$ less memory consumption. 
Overall, our results demonstrate that \emph{SheetReader} provides runtime and memory-efficient spreadsheet parsing.

\begin{figure}
  \centering
  \includegraphics[width=1.01\linewidth, keepaspectratio]{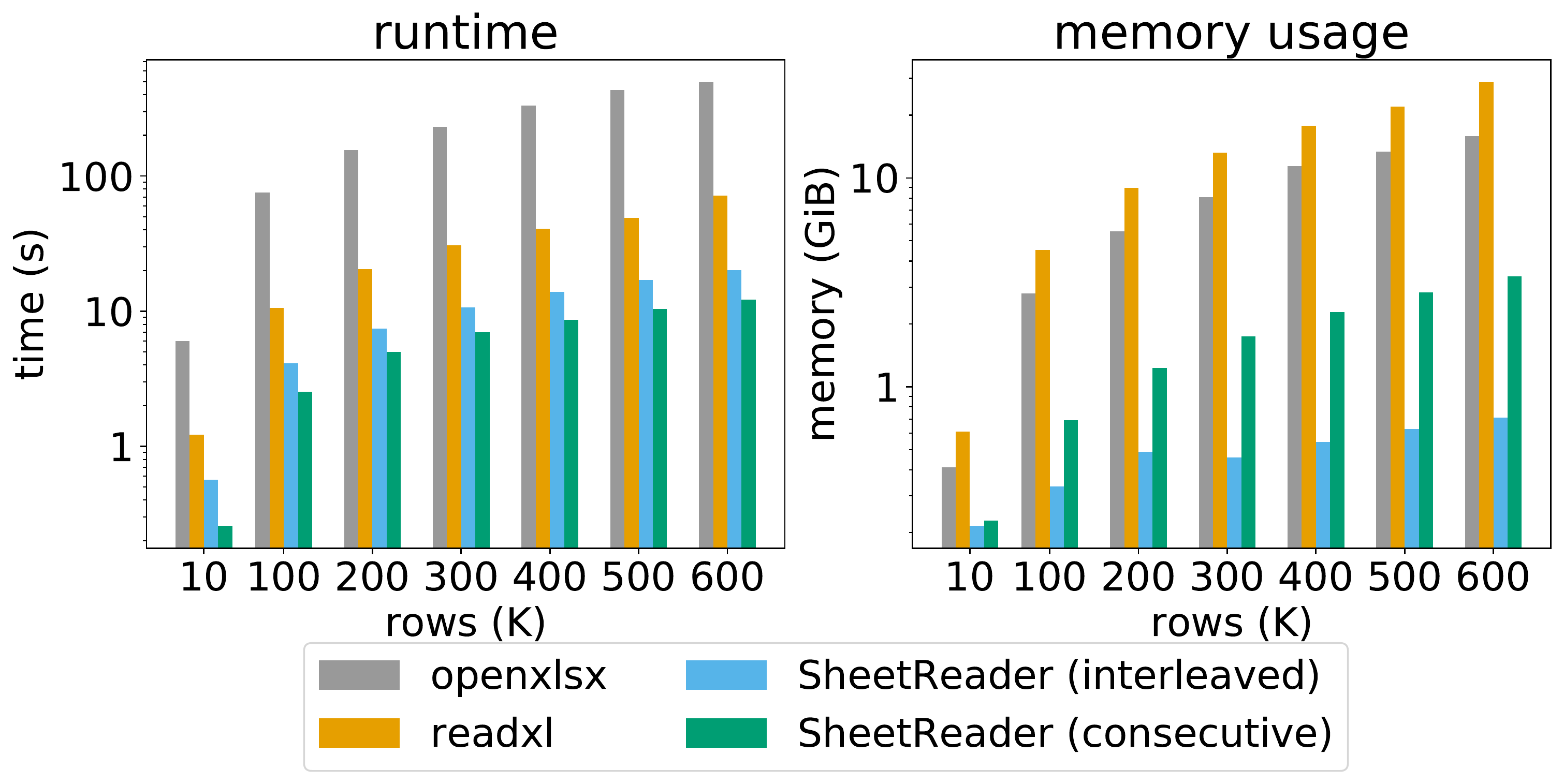}
  \caption{Performance overview of \emph{SheetReader} and existing R packages for parsing synthetic spreadsheets containing only numeric values.}
  \label{fig:evaluation_overview}
\end{figure}

{\ParHead Scalability with Spreadsheet Size.} \Cref{fig:evaluation_overview} shows the runtime and memory usage as we increase the size of our synthetically generated spreadsheets. 
Comparing the runtime performance, \texttt{openxlsx} exhibits very long runtimes even for moderately sized files, taking more than 2 minutes for a spreadsheet with 200,000 rows.
\texttt{Readxl}, which is the fastest existing solution for loading spreadsheets into R, reaches 65 seconds for the largest file. 
Our approach, \emph{SheetReader}, outperforms both baselines by around 2.5 to 3 times across all tested worksheet sizes.

In terms of memory efficiency, \emph{SheetReader} has a considerable lead over the other packages, consuming at most 728 MB for the largest file of 600,000 rows. 
Specifically, \emph{SheetReader} consumes up to $40\times$ and $20\times$ less memory than \texttt{readxl} (29.5 GB) and \texttt{openxlsx} (16.3 GB), respectively.

The excessive memory usage of \texttt{readxl} is caused by its underlying XML DOM parsing approach.
The generated DOM tree that is kept in memory for subsequent processing consumes large amounts of memory.
As a result, the memory usage of \texttt{readxl} is consistently over 10 times more than the size of the uncompressed source worksheet, reaching almost 30 GB for 600,000 rows.  
For consumer machines, even worksheets with 200,000 or 300,000 rows can saturate all available memory (9 GB and 13.5 GB respectively in this benchmark), which would in turn also impact the runtime.
That is, the runtime measurements will become significantly worse than the ones shown here if we use a machine that does not have a sufficient amount of memory. 

The package \texttt{openxlsx} employs an approach that can be considered a mix between DOM and SAX parsing. 
Instead of extracting the whole document into a DOM tree, it extracts only the significant parts of the document into lists using regular expressions.
However, it does not directly process the extracted values.
Specifically, while the extraction of values from the worksheet is done in C\texttt{++}, the lists containing the values are returned to R. 
Then, the R wrapper function that wraps the extraction processes these values further to build the target \texttt{Dataframe}.
Overall, while \texttt{openxlsx} uses considerably less memory than \texttt{readxl}, its memory usage is still excessive, i.e., around 8 GB for 300,000 rows and reaching 16 GB for 600,000 rows.

\subsection{SheetReader Analysis}

\begin{figure}
  \centering
  \includegraphics[width=\linewidth]{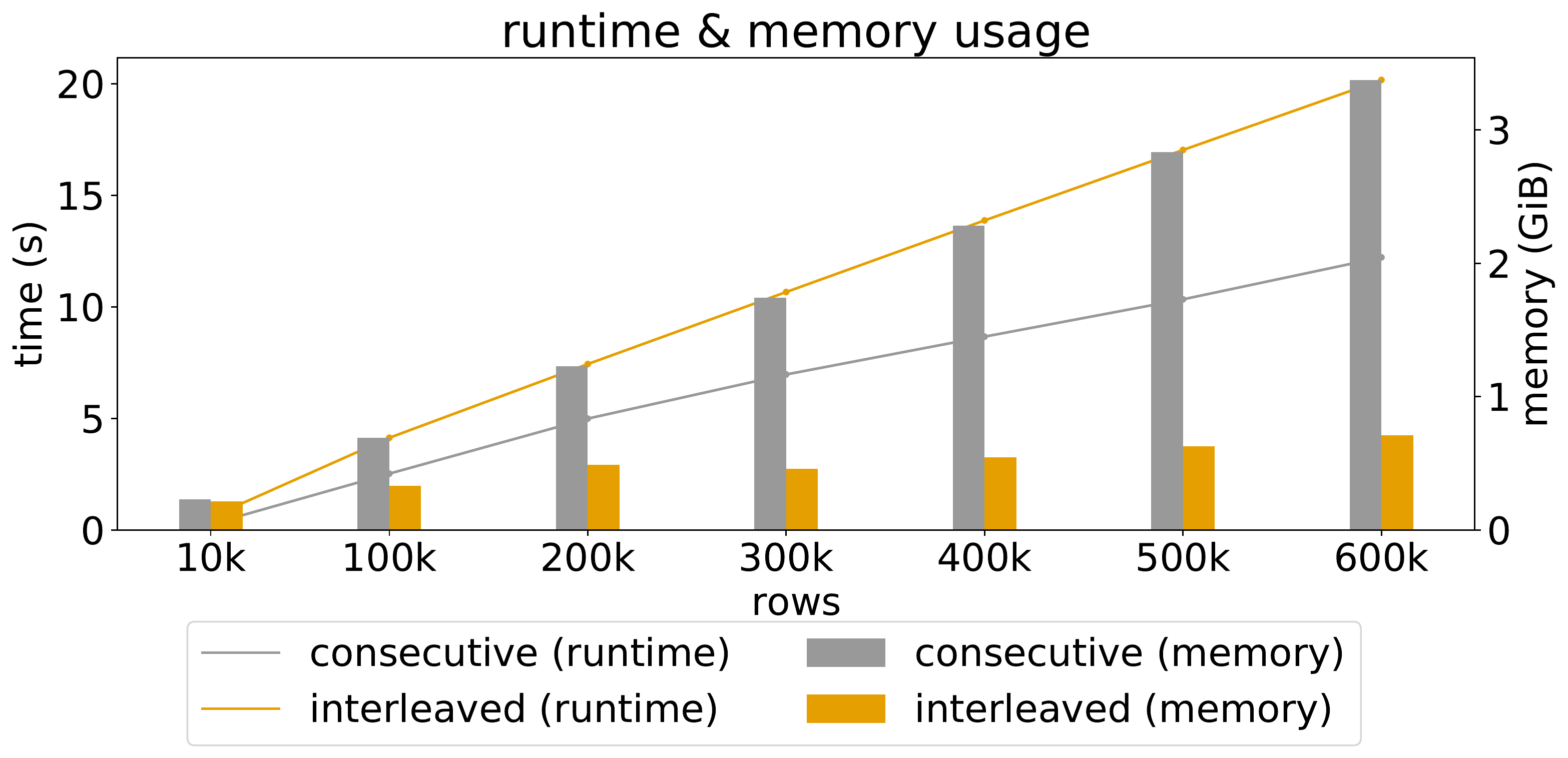}
  \caption{Benchmarks of the \emph{Consecutive} and the \emph{Interleaved} parsing approaches.}
  \label{fig:evaluation_sequential_vs_interleaved}
\end{figure}

{\ParHead Parsing Approaches Comparison.} We introduced two different parsing approaches for \emph{SheetReader}; \emph{consecutive} and \emph{interleaved}.
This benchmark studies the trade-offs between their runtime and memory usage.
Particularly, it aims to determine the speedup of the \emph{consecutive} over the \emph{interleaved} approach, and to show how the \emph{consecutive} approach achieves this speedup at the expense of an increased memory usage.

\Cref{fig:evaluation_sequential_vs_interleaved} shows the results when applying both approaches to the same synthetic spreadsheets and increasing the spreadsheet size.
Both approaches exhibit a linear increase in runtime and memory usage that is proportional to the size, with the \emph{consecutive} approach consistently having a better runtime but also substantially higher memory usage.
In contrast, the increase in memory usage of the \emph{interleaved} approach is negligible. 

In the \emph{consecutive} approach, the decompression step requires two buffers, one for the compressed and one for the decompressed content.
Therefore, the maximum memory usage is effectively determined by the sum of the sizes of the compressed and the decompressed worksheet.
The intermediate data structure is allocated only after the deallocation of the compressed document (i.e., after decompression), while it is generally considerably smaller than the worksheet. 
As such, it has no impact on the maximum memory usage.
In contrast, in the \emph{interleaved} approach, since the actual parsing process uses a constant amount of memory, any increase of the memory usage over different worksheet sizes is caused by the intermediate data structure, whose size depends on the input worksheet.

Furthermore, the benchmark shows that while the runtime rises linearly for both approaches, the increase for the \emph{interleaved} approach is stronger than for the \emph{consecutive} one, culminating in a difference of around 8 seconds for 600,000 rows.

Our benchmark confirms the advantages and disadvantages of both parsing approaches discussed in Section~\ref{sec:approach}. 
Additionally, based on the experimental results, we propose using the \emph{interleaved} approach as the "safe default" option because it loads the spreadsheet data in an acceptable amount of time while only rarely consuming more memory than the one that is already required by the target environment to store the same data.
Users can choose the \emph{consecutive} approach if they require faster loading times and have a machine with a sufficient amount of memory.

\begin{figure}
  \centering
  \includegraphics[width=1.01\linewidth, keepaspectratio]{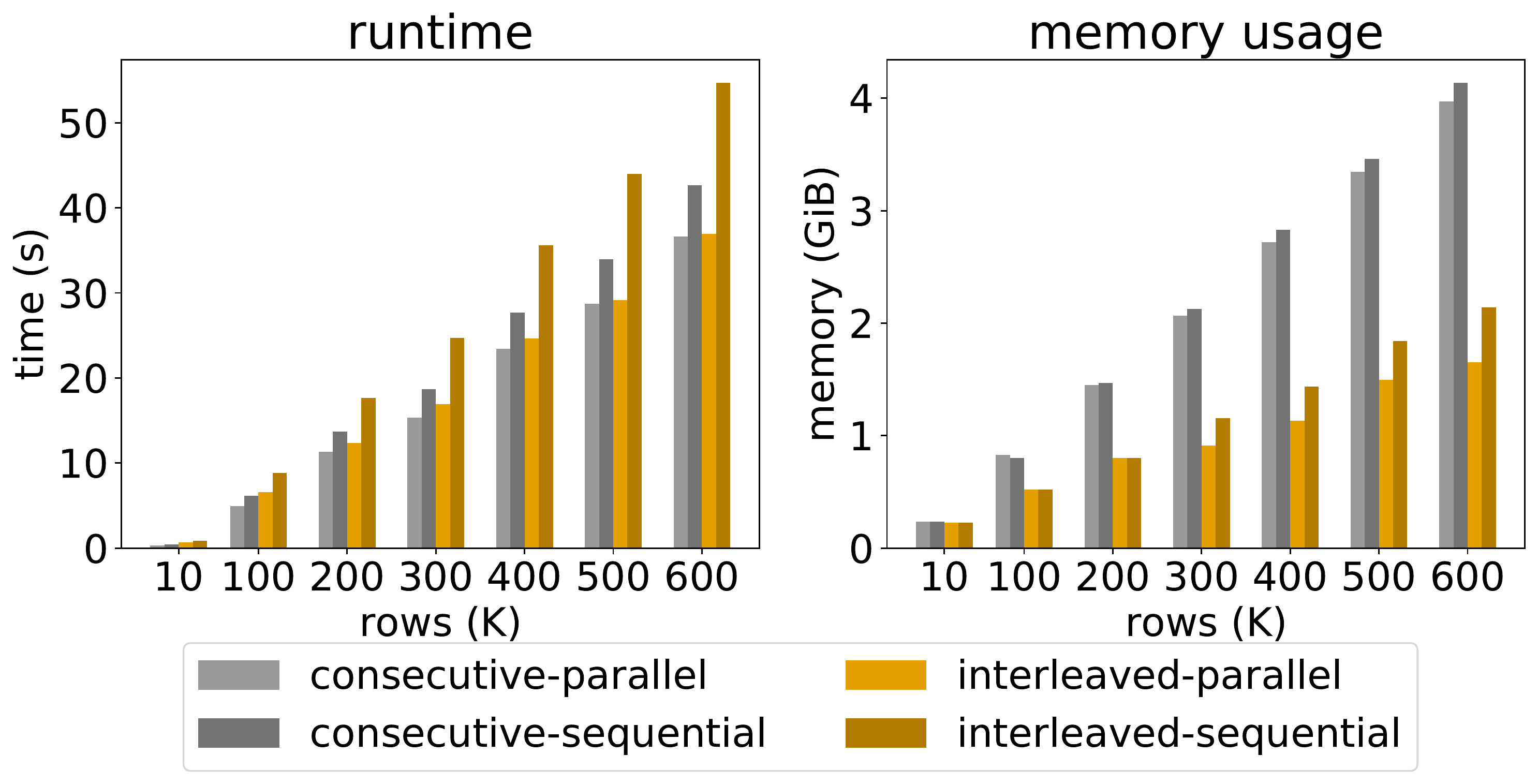}
  \caption{Benchmarks of parsing the shared strings sequentially or parallel to the worksheet.}
  \label{fig:strings_sequential_vs_parallel}
\end{figure}

{\ParHead Parallelizing Worksheet and Shared Strings Parsing.} Apart from the choice of parsing approach, in the case of spreadsheet systems that store strings separately from the worksheet such as Excel, we can also choose whether to parse these documents sequentially or in parallel. 
To compare the performance of these two approaches, we generate synthetic spreadsheets for various row counts that contain a mix of different data types.
Specifically, the synthetically generated mixed-type spreadsheets have 40 columns of floating point values, 30 columns of integer values, 20 columns of text with 25\% unique values, and 10 columns of text with 75\% unique values. 

As expected, \Cref{fig:strings_sequential_vs_parallel} shows that parsing the shared strings and the worksheet in parallel yields runtime improvements.
The \emph{interleaved} parsing approach benefits the most from this parallelization, reaching a runtime reduction of around 30\%. 
However, contrary to our expectations, the parallel approach has a lower memory usage than the sequential one for almost all benchmarks. 
To determine the cause of this, next we examine the memory characteristics of the benchmarks in more detail.

\begin{figure}
  \centering
  \includegraphics[width=1.01\linewidth, keepaspectratio]{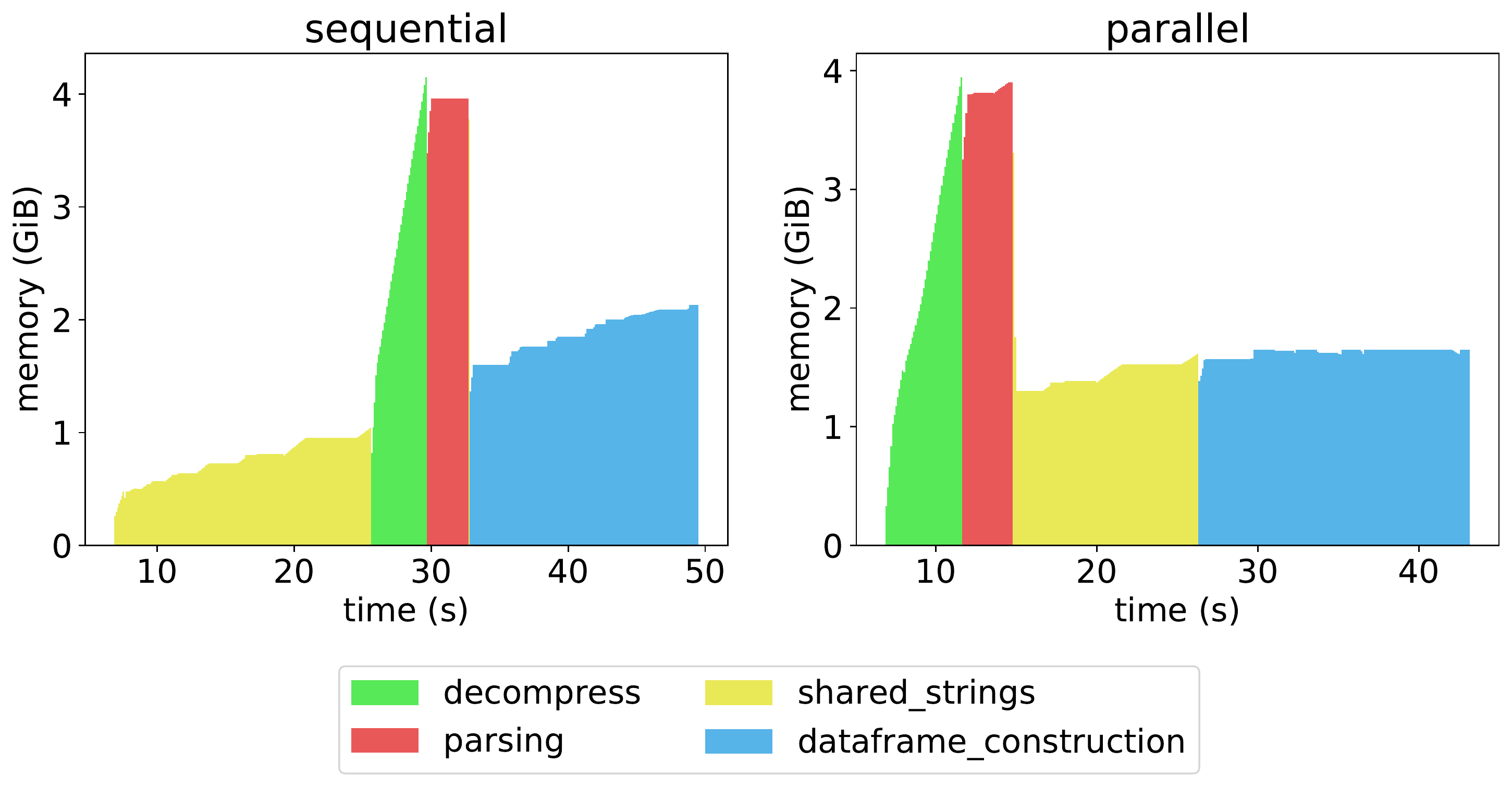}
  \caption{Memory measurements of the sequential and the parallel approach (\emph{consecutive} parsing).}
  \label{fig:sequential_vs_interleaved_time_memory}
\end{figure}

{\ParHead Memory Usage Analysis.} \Cref{fig:sequential_vs_interleaved_time_memory} shows a detailed memory profile of the sequential and parallel approaches when parsing the largest document (600,000 rows) from our previous experiment using the \emph{consecutive} approach.
To identify when and where the maximum memory usage occurred, we measure it periodically and associate different time spans with different steps in the parsing process. 
The green \texttt{decompress} and red \texttt{parsing} sections correspond to worksheet parsing, while the yellow \texttt{shared\_strings} section combines both steps in the shared strings parsing process. 
In the parallel benchmark, the yellow section depicts the extra time taken by shared strings parsing. 
While worksheet parsing finishes at around 15 seconds, shared strings parsing takes over 10 extra seconds to finish, delaying the dataframe construction.

Both benchmarks show that parsing the shared strings table is two to three times slower than parsing the worksheet.
The runtime difference can be explained by the inability to parallelize the parsing process for the shared strings table, while the worksheet is distributed among 8 threads.
Furthermore, the memory usage increases steadily during the parsing of the shared strings table.
This increase in memory usage stems from allocating space to copy the strings out of the original document, so that we can return them to the user after the deallocation of the document.

The reason why parsing the worksheet and shared strings table in parallel has a lower memory usage than doing so sequentially is a combination of three factors: the dynamic string allocations, the long runtime of shared strings parsing compared to worksheet parsing, and the order of the two parsing steps in the sequential approach.
The sequential approach processes all shared strings and allocates space for them before decompressing the worksheet, which represents a constant base memory usage for all subsequent steps, including any processing of the worksheet where shared strings are not required. 
As a result, the maximum memory usage is reached when the worksheet is decompressed, since the copied strings occupy additional memory on top of the decompressed content.  
In the parallel approach, since the shared strings parsing step is slow, by the time all strings are copied, the worksheet is fully parsed and the source document has been deallocated.

We conclude that for the sequential approach, the parsing of the shared strings table should occur after the worksheet parsing to reduce the maximum memory usage.
Parsing the strings after the worksheet has the additional benefit of allowing to filter out unneeded strings, i.e., strings that do not occur in the specified sheet.
Swapping the order of the parsing steps in our prototype is straightforward, as these steps are independent. 

\begin{figure}
  \centering
  \includegraphics[width=\linewidth, keepaspectratio]{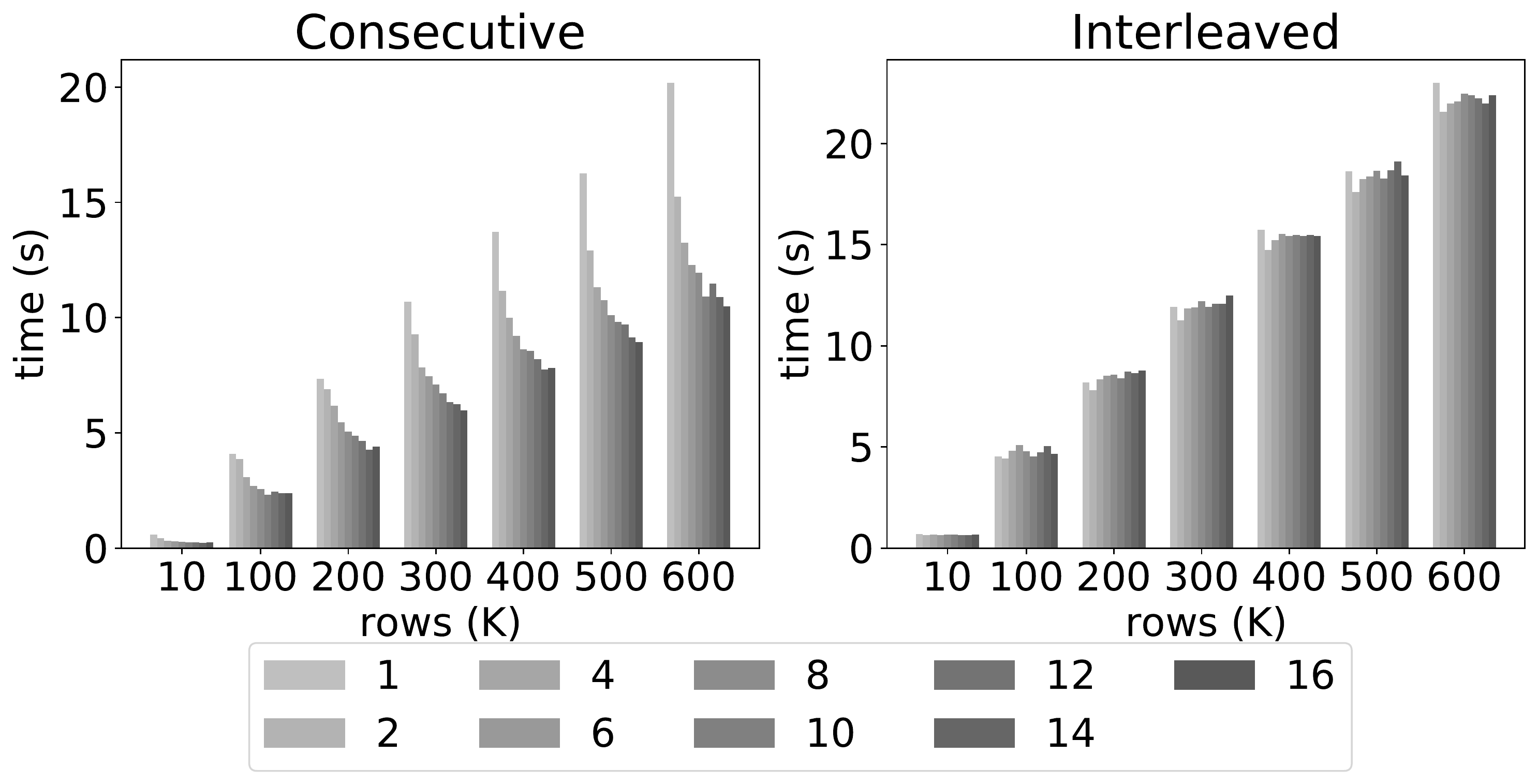}
  \caption{Impact of the number of threads.}
  \label{fig:evaluation_interleaved_threads}
\end{figure}

{\ParHead Impact of Thread Count.} To evaluate the effectiveness of our parallelization efforts, we measure the impact of the number of used threads on the runtime for both the \emph{consecutive} and the \emph{interleaved} approach. 
\Cref{fig:evaluation_interleaved_threads} shows that the benefits decrease as we increase the thread count in both parsing approaches.
Particularly for the \emph{interleaved} approach, any noticeable runtime improvement (5 to 10\%) stops at only two parsing threads, while the runtime actually increases with more than two threads.
Further analysis when running the benchmarks reveals that the decompression thread becomes the limiting factor at this point, so that any additional parsing threads only introduce more synchronization overhead.
Regardless of the number of parsing threads, the decompression is too slow and results in idle threads waiting for a new available buffer element. 
Thus, the only way to further reduce the runtime is accelerating the decompression.

The \emph{consecutive} approach exhibits a more gradual runtime reduction, with the increase from 1 to 8 threads reducing the runtime by almost half (20 to 12 seconds for 600,000 rows), while the increase from 8 to 16 threads only has a marginal impact (12 to 10.5 seconds). 
We are again effectively limited by the speed of the decompression step. 
Since this approach performs parsing after the completion of the decompression, the slow decompression component imposes the lower limit for the runtime.

\subsection{Parallel Decompression}

\begin{figure}
  \centering
  \includegraphics[width=0.9\linewidth, keepaspectratio]{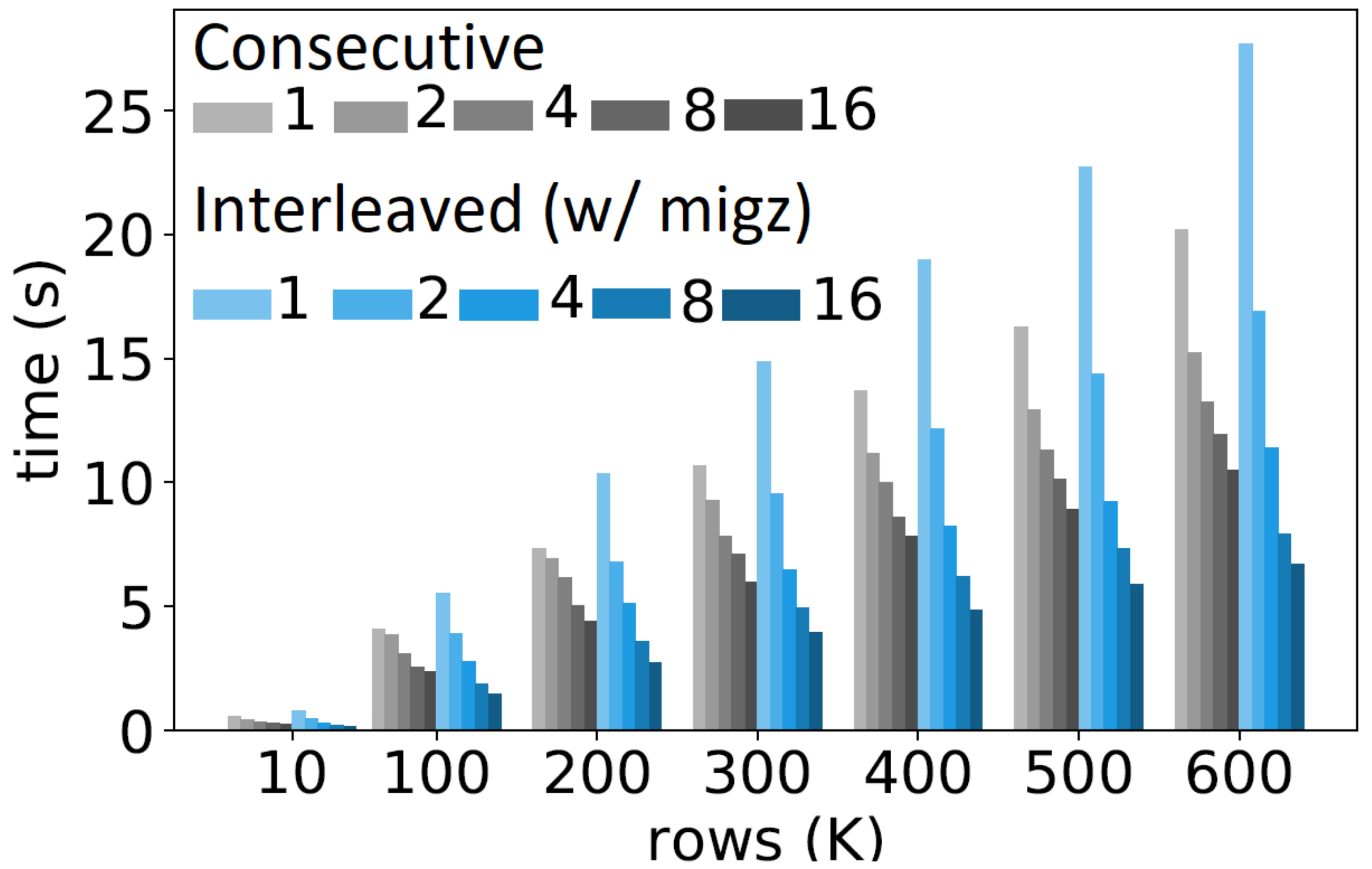}
  \caption{Impact of parallel decompression.}
  \label{fig:evaluation_parallel_decompression}
  \vspace{-0.1cm}
\end{figure}

Since decompression is a runtime bottleneck, we performed an experiment to determine the advantage that we can get from parallelizing it.  
To that end, and since the current compression used by the OOXML and ODF formats does not support parallel decompression, we extracted the worksheet XML files from the Excel files and re-compressed them with a modified Deflate algorithm based on the MiGz library\footnote{\url{https://github.com/linkedin/migz}}.
Furthermore, we established boundaries in the Deflate stream after which there are no back-references to previous blocks and stored the offsets of these boundaries in the file metadata. 
The result is a valid Deflate stream that can be decompressed with any existing library.
A decompression algorithm can now start full decompression of the stream from any of the boundaries without requiring to first fully decompress the previous blocks in the stream. 
Therefore, we can parallelize the decompression of a single document. 
Specifically, in our implementation we assign separate threads to equally spaced boundaries.
Each thread performs decompression and parsing in an interleaved manner (i.e., using our \emph{interleaved} approach) until it reaches the next boundary.

\Cref{fig:evaluation_parallel_decompression} compares the \emph{consecutive} approach without parallel decompression with the \emph{interleaved} approach that parallelizes the decompression using our MiGz-derived algorithm when increasing the thread count.  
We see that the parallel decompression implementation outperforms the \emph{consecutive} approach when using more than 2 threads, especially for larger files. 
In most cases, using only 4 threads, the parallel decompression implementation achieves the same runtime as the consecutive approach with 16 threads. 
In turn, 16 threads enable the MiGz-derived algorithm to lower the runtime by an additional 35\%. 
We also observe that increasing the number of threads has a larger effect on the runtime for the fully parallel implementation.
Finally, we note that since the individual threads employ the \emph{interleaved} parsing approach, the memory usage is significantly lower than the one of the \emph{consecutive} approach.
Therefore, parallel decompression allows us to further reduce the runtime while retaining low memory usage.

\subsection{Summary}
Our experimental comparison with the existing solutions shows the efficiency of our proposed spreasheet parsing architecture. 
Overall, \emph{SheetReader} with \emph{interleaved} parsing loads spreadsheet files $2\times$ to $3\times$ faster than the fastest existing solution while consuming up to $20\times$ less memory than the most memory-efficient existing solution. 
As such, our parser can process large spreadsheets on current consumer machines without requiring an excessive amount of resources and degrading the user experience.

\emph{SheetReader} offers an alternative consecutive parsing approach which reduces the runtime by an additional $40\%$ but also increases the memory usage by more than $4$ times. 
Furthermore, parsing the shared strings and the worksheet in parallel reduces the runtime by $15\%$ and $30\%$ when using consecutive and interleaved parsing, respectively.
Parallelizing the parsing process itself grants the \emph{consecutive} parsing approach a $20\%$ to $30\%$ runtime reduction when using 4 threads, while having only negligible impact on the \emph{interleaved} approach. 
Finally, we show that we could achieve significant additional performance improvements by parallelizing the decompression.
Unfortunately, the current specification of the OOXML and ODF formats does not allow this parallelization.

\section{Related Work}
\label{sec:related_work}
While there is some work on extracting specific content from spreadsheets, e.g., tables~\cite{table_identification}, there is no related work directly focusing on spreadsheet parsing; current approaches rely on generalized XML parsing.
Therefore, we review techniques proposed for efficient parsing of XML and other text-based formats.

{\ParHead XML Parser Parallelization.}
Parallelizing XML parsing is a non-trivial task~\cite{nicola2003xml}.
As the XML format is self-describing, the difficulty lies in splitting an XML document into chunks that can be parsed in parallel.
One line of work proposes a two-pass approach to build an XML skeleton structure, which allows to split the document before parsing and merge the individual results efficiently~\cite{lu2006parallel,pan2007static}.
Follow-up work proposes to also parallelize the first pass by letting multiple threads create multiple skeletons for each chunk, and then merging them into one~\cite{pan2008transducers}.
Another line of work shows that producing chunks with an arbitrary number of start and end XML tags and then merging partial results, offers better scalability than the two-step approach on multicore systems~\cite{shah2009a}.
Furthermore, leveraging SIMD (single-instruction multiple-data) instructions of modern CPUs, allows to  parallelize character scanning and to avoid cache misses, conditional branches and branch mispredictions, thereby further minimizing the parsing runtime~\cite{cameron2008high}.
Both lines of work are complementary to SheetReader, as they can be employed to better split the XML and parallelize character scanning at a lower level.

{\ParHead XML Parser Compilers.}
An approach to accelerate the parsing procedure is to specialize the parser to a given schema.
Parser compilers generate parsers based on predefined schemata.
XML Screamer~\cite{kostoulas2006xml} compiles specialized parsers that merge parsing and deserialization to avoid expensive data copying and transformation operations.
Chiu and Lu~\cite{chiu2004compiler} propose an intermediate representation with a generalized automata approach, through which they generate efficient parsers.
Although schema-based specialization leads to better performance, it does not directly exploit spreadsheet-specific properties to further optimize parsing.

{\ParHead Parsing Text-Based Formats.}
Parsing widespread text-based formats, e.g., CSV and JSON, is similarly challenging as parsing XML.
There has been extensive work on improving the performance of CSV parsing, e.g., by employing speculative parsing techniques~\cite{ge2019speculative}, by optimizing the parsing process for multicore CPUs~\cite{muhlbauer2013instant}, and by employing GPUs for parallelization~\cite{stehle2019parparaw, kumaigorodski2021fast}.
In-situ data processing approaches also employ several optimizations, such as selective parsing and just-in-time compilation~\cite{karpathiotakis2014adaptive}. %
Furthermore, multi-hypothesis CSV parsing addresses the challenge of validating files with unknown schemata~\cite{dohmen2017multi}.

The JSON format shares more similarities with the XML format, as they are both self-describing.
Several approaches have been proposed to improve the performance of JSON parsing.
For example, Sparser~\cite{palkar2018filter} employs raw filtering through SIMD instructions before parsing, while Mison~\cite{li2017mison} speculatively predicts the physical location of necessary fields through structural indices.
Moreover, simdjson~\cite{langdale2019parsing} proposes to limit the set of employed instructions to increase the parsing and validation performance of JSON documents on commodity CPUs.
We see these lines of work as orthogonal to ours, as they can be applied in the context of SheetReader to further increase performance.

\section{Conclusions}
\label{sec:conclusion}
Spreadsheet systems are popular for accessible data analysis but have limited capabilities when it comes to data science applications. 
Existing solutions for loading spreadsheets into other data science environments to perform advanced analytics exhibit critical performance problems in terms of either runtime or memory usage.
To address these problems, this paper introduces \texttt{SheetReader}, a specialized spreadsheet parsing architecture that operates in two different parsing modes, \emph{consecutive} and \emph{interleaved}.
To improve the runtime, \texttt{SheetReader} parallelizes the parsing by exploiting the flat and repeating structures inherently found in spreadsheet formats. 
It further uses task parallelism to process worksheets and strings of the spreadsheet concurrently.
To reduce the memory utilization, \texttt{SheetReader} tightly couples decompression and parsing.
To provide a general solution for different target environments, it stores the retrieved spreadsheet values in an environment-agnostic intermediate data structure. 
That way, one can easily create bindings for different targets without the need to modify the core parser.

Our evaluation showed that \texttt{SheetReader} is highly efficient in terms of both runtime and memory usage.
The \emph{consecutive} approach offers a significant improvement in runtime and a moderate reduction in memory usage, while the \emph{interleaved} approach yields a more moderate runtime improvement but has very low memory consumption.
Since decompression creates a bottleneck, we also introduced and evaluated a method for parallel decompression, showing that with fully data-parallel processing we can further reduce the runtime while keeping the memory usage low.

In future work, we plan to investigate the applicability of existing solutions that partially parallelize the decompression of general Deflate streams, such as pugz~\cite{pugz}.
Furthermore, we plan to extend our prototype for other data science environments (e.g., Python) and to incorporate SheetReader as a DBMS spreadsheet wrapper, similar to SCANRAW~\cite{cheng_parallel_2014}.

\begin{acks}
This work was funded by the German Ministry for Education \& Research as BIFOLD - Berlin Institute for the Foundations of Learning \& Data (ref. 01IS18025A and 01IS18037A).
We would like to thank Sergey Redyuk for his assistance in setting up the benchmarking environment and Clemens Lutz for his feedback.
\end{acks}

\bibliographystyle{abbrv}
\bibliography{bibliography}

\begin{thebibliography}{10}

\bibitem{ooxml}
{Standard ECMA-376 Office Open XML File Formats}.
\newblock
  \url{https://www.ecma-international.org/publications/standards/Ecma-376.htm}.
\newblock Accessed on 2021-12-14.

\bibitem{cameron2008high}
R.~D. {Cameron}, K.~S. {Herdy}, and D.~{Lin}.
\newblock {High performance XML parsing using parallel bit stream technology}.
\newblock In {\em CASCON}, pages 222--235, 2008.

\bibitem{cheng_parallel_2014}
Y.~Cheng and F.~Rusu.
\newblock Parallel in-situ data processing with speculative loading.
\newblock In {\em {SIGMOD}}, pages 1287--1298, 2014.

\bibitem{chiu2004compiler}
K.~Chiu and W.~Lu.
\newblock {A compiler-based approach to schema-specific XML parsing}.
\newblock In {\em 1st Int'l. Workshop on High Performance XML Processing},
  2004.

\bibitem{deflate}
P.~Deutsch.
\newblock Rfc1951: Deflate compressed data format specification version 1.3,
  1996.

\bibitem{dohmen2017multi}
T.~D{\"o}hmen, H.~M{\"u}hleisen, and P.~Boncz.
\newblock Multi-hypothesis {CSV} parsing.
\newblock In {\em SSDBM}, pages 1--12, 2017.

\bibitem{ge2019speculative}
C.~Ge, Y.~Li, E.~Eilebrecht, B.~Chandramouli, and D.~Kossmann.
\newblock Speculative distributed {CSV} data parsing for big data analytics.
\newblock In {\em SIGMOD}, pages 883--899, 2019.

\bibitem{karpathiotakis2014adaptive}
M.~Karpathiotakis, M.~Branco, I.~Alagiannis, and A.~Ailamaki.
\newblock Adaptive query processing on raw data.
\newblock {\em PVLDB}, 7(12):1119--1130, 2014.

\bibitem{pugz}
M.~Kerbiriou and R.~Chikhi.
\newblock {Parallel Decompression of Gzip-Compressed Files and Random Access to
  DNA Sequences}.
\newblock In {\em IPDPSW}, pages 209--217, 2019.

\bibitem{table_identification}
E.~Koci, M.~Thiele, O.~Romero, and W.~Lehner.
\newblock Table identification and reconstruction in spreadsheets.
\newblock In {\em CAiSE}, pages 527--541, 2017.

\bibitem{kostoulas2006xml}
M.~G. Kostoulas, M.~Matsa, N.~Mendelsohn, E.~Perkins, A.~Heifets, and
  M.~Mercaldi.
\newblock {{XML} screamer: an integrated approach to high performance XML
  parsing, validation and deserialization}.
\newblock In {\em WWW}, pages 93--102, 2006.

\bibitem{kumaigorodski2021fast}
A.~Kumaigorodski, C.~Lutz, and V.~Markl.
\newblock {Fast CSV Loading Using GPUs and RDMA for In-Memory Data Processing}.
\newblock {\em BTW}, 2021.

\bibitem{parsing_survey}
T.~C. Lam, J.~J. Ding, and J.-C. Liu.
\newblock {XML Document Parsing: Operational and Performance Characteristics}.
\newblock {\em Computer}, 41(9):30--37, 2008.

\bibitem{langdale2019parsing}
G.~Langdale and D.~Lemire.
\newblock Parsing gigabytes of {JSON} per second.
\newblock {\em The VLDB Journal}, 28(6):941--960, 2019.

\bibitem{Li2009}
C.~Li.
\newblock {\em XML Parsing, SAX/DOM}, pages 3598--3601.
\newblock Springer US, 2009.

\bibitem{li2017mison}
Y.~Li, N.~R. Katsipoulakis, B.~Chandramouli, J.~Goldstein, and D.~Kossmann.
\newblock {Mison: a fast JSON parser for data analytics}.
\newblock {\em PVLDB}, 10(10):1118--1129, 2017.

\bibitem{lu2006parallel}
W.~Lu, K.~Chiu, and Y.~Pan.
\newblock {A parallel approach to XML parsing}.
\newblock In {\em GridCom}, pages 223--230, 2006.

\bibitem{muhlbauer2013instant}
T.~M{\"u}hlbauer, W.~R{\"o}diger, R.~Seilbeck, A.~Reiser, A.~Kemper, and
  T.~Neumann.
\newblock {Instant loading for main memory databases}.
\newblock {\em PVLDB}, 6(14):1702--1713, 2013.

\bibitem{nicola2003xml}
M.~Nicola and J.~John.
\newblock {{XML} parsing: a threat to database performance}.
\newblock In {\em CIKM}, pages 175--178, 2003.

\bibitem{palkar2018filter}
S.~Palkar, F.~Abuzaid, P.~Bailis, and M.~Zaharia.
\newblock {{Filter before you parse: Faster analytics on raw data with
  sparser}}.
\newblock {\em PVLDB}, 11(11):1576--1589, 2018.

\bibitem{pan2007static}
Y.~Pan, W.~Lu, Y.~Zhang, and K.~Chiu.
\newblock {A static load-balancing scheme for parallel XML parsing on multicore
  CPUs}.
\newblock In {\em CCGrid}, pages 351--362, 2007.

\bibitem{pan2008transducers}
Y.~Pan, Y.~Zhang, and K.~Chiu.
\newblock {Simultaneous transducers for data-parallel XML parsing}.
\newblock In {\em IPDPS}, pages 1--12, 2008.

\bibitem{rahman2021noah}
S.~{Rahman}, M.~{Bendre}, Y.~{Liu}, S.~{Zhu}, Z.~{Su}, K.~{Karahalios}, and
  A.~G. {Parameswaran}.
\newblock {NOAH: interactive spreadsheet exploration with dynamic hierarchical
  overviews}.
\newblock {\em PVLDB}, 14(6):970--983, 2021.

\bibitem{rahman2020benchmarking}
S.~Rahman, K.~Mack, M.~Bendre, R.~Zhang, K.~Karahalios, and A.~Parameswaran.
\newblock {Benchmarking Spreadsheet Systems}.
\newblock In {\em SIGMOD}, pages 1589--1599, 2020.

\bibitem{shah2009a}
B.~{Shah}, P.~R. {Rao}, B.~{Moon}, and M.~{Rajagopalan}.
\newblock {A Data Parallel Algorithm for XML DOM Parsing}.
\newblock In {\em XSym}, volume 5679, pages 75--90, 2009.

\bibitem{stehle2019parparaw}
E.~Stehle and H.~Jacobsen.
\newblock {ParPaRaw: Massively Parallel Parsing of Delimiter-Separated Raw
  Data}.
\newblock {\em {PVLDB}}, 13(5):616--628, 2020.

\end{thebibliography}

\end{document}